\documentstyle[12pt,epsf,epsfig,amsmath]{article}
\newfont{\thiplo}{msbm10 scaled\magstep 2}
\newfont{\gothic}{eufb10 scaled\magstep 2}
\newfont{\unc}{eurb10} 
\newskip\humongous \humongous=0pt plus 1000pt minus 1000pt
\def\caja{\mathsurround=0pt}\def\eqalign#1{\,\vcenter{\openup1\jot \caja
        \ialign{\strut \hfil$\displaystyle{##}$&$
        \displaystyle{{}##}$\hfil\crcr#1\crcr}}\,}
\newif\ifdtup

\def\eqright #1\cr{\noalign{\hfill$\displaystyle{{}#1}$}}
\def\eqleft #1\cr{\noalign{\noindent$\displaystyle{{}#1}$\hfill}}

\def\oldreffmt#1{\rlap{[#1]} \hbox to 2\parindent{}}

\def\figfmt#1{\rlap{Figure {#1}} \hbox to 1in{}}

\def\sectioneq{\def\theequation{\thesection.\arabic{equation}}{\let
\holdsection=\section\def\section{\setcounter{equation}{0}\holdsection}}}%

\newcounter{holdequation}

\def\begineq #1\endeq{$$ \refstepcounter{equation}\eqalign{#1}\eqno
	(\theequation) $$}
\def\contlimit{\,{\hbox{$\longrightarrow$}\kern-1.8em\lower1ex
\hbox{${\scriptstyle (a\rightarrow0)}$}}\,}
\def\centeron#1#2{{\setbox0=\hbox{#1}\setbox1=\hbox{#2}\ifdim
\wd1>\wd0\kern.5\wd1\kern-.5\wd0\fi
\copy0\kern-.5\wd0\kern-.5\wd1\copy1\ifdim\wd0>\wd1
\kern.5\wd0\kern-.5\wd1\fi}}
\def\centerover#1#2{\centeron{#1}{\setbox0=\hbox{#1}\setbox
1=\hbox{#2}\raise\ht0\hbox{\raise\dp1\hbox{\copy1}}}}
\def\centerunder#1#2{\centeron{#1}{\setbox0=\hbox{#1}\setbox
1=\hbox{#2}\lower\dp0\hbox{\lower\ht1\hbox{\copy1}}}}
\def\lsim{\;\centeron{\raise.35ex\hbox{$<$}}{\lower.65ex\hbox
{$\sim$}}\;}
\def\gsim{\;\centeron{\raise.35ex\hbox{$>$}}{\lower.65ex\hbox
{$\sim$}}\;}

\def\super#1{\ifmmode \hbox{\textsuper{#1}}\else\textsuper{#1}\fi}
\def\textsuper#1{\newcount\holdspacefactor\holdspacefactor=\spacefactor
$^{#1}$\spacefactor=\holdspacefactor}

\def\getcite#1,{\advance\citenumber by1
\def\getcitearg{#1}\def\lastarg{@}
\ifnum\citenumber=1
\ref{#1}\let\next=\getcite\else\ifx\getcitearg\lastarg\let\next=\relax
\else ,\ref{#1}\let\next=\getcite\fi\fi\next}

\def\pom{{\rm P\kern -0.53em\llap I\,}}
\def\spom{{\rm P\kern -0.36em\llap \small I\,}}
\def\sspom{{\rm P\kern -0.33em\llap \footnotesize I\,}}

\relax
\def\contlimit{\,{\hbox{$\longrightarrow$}\kern-1.8em\lower1ex
\hbox{${\scriptstyle (a\rightarrow0)}$}}\,}
\def\upon #1/#2 {{\textstyle{#1\over #2}}}
\relax
\renewcommand{\thefootnote}{\fnsymbol{footnote}} 

\def\mainhead#1{\setcounter{equation}{0}\addtocounter{section}{1}
  \vbox{\begin{center}\large\bf #1\end{center}}\nobreak\par}
\sectioneq

\def\til#1{\centeron{\hbox{$#1$}}{\lower 2ex\hbox{$\char'176$}}}
\def\tild#1{\centeron{\hbox{$\,#1$}}{\lower 2.5ex\hbox{$\char'176$}}}
\def\sumtil{\centeron{\hbox{$\displaystyle\sum$}}{\lower
-1.5ex\hbox{$\widetilde{\phantom{xx}}$}}}

\begin{document} 

\begin{titlepage} 

\rightline{\vbox{\halign{&#\hfil\cr
&\today\cr}}} 
\vspace{0.25in} 

\begin{center} 
  
{\large\bf The LHC Pomeron and Unification of the Standard 
Model\footnote{Presented 
at the Small-x and Diffraction Workshop, Fermilab,March 2007.}  
\newline $~$
\newline {\it - a Bound-State S-Matrix Within a Fixed-Point Field Theory ?}}

\medskip

Alan. R. White\footnote{arw@hep.anl.gov } 

\vskip 0.6cm
 
\centerline{Argonne National Laboratory}
\centerline{9700 South Cass, Il 60439, USA.}
\vspace{0.5cm}

\end{center}

\begin{abstract} 
 
The Critical Pomeron solution of 
high-energy unitarity leads to a unique underlying massless field theory
that might be the origin of the Standard Model. A
color sextet quark sector - producing both
electroweak symmetry breaking and dark matter - is added to QCD to saturate
asymptotic freedom. The sextet sector is then embedded uniquely in ``QUD'' - 
an anomaly free, 
just asymptotically free,  
massless SU(5) theory with elementary lepton and triplet quark sectors very close
to the Standard Model. A multi-regge bound-state S-Matrix is constructed
using infra-red divergent scaling reggeon interactions that  
couple via massless fermion chiral anomalies. 
Within the QCD sub-sector  
there is an ``anomalous wee gluon'' critical phenomenon that 
produces a spectrum with confinement and chiral symmetry breaking. 
The exponentiation of left-handed gauge boson divergences
implies that the full set of composite interactions and 
the low-mass spectrum of QUD could be just those of the Standard Model. 
All particles, including neutrinos,
appear as massive, Goldstone boson related, bound-states and there is no 
Higgs field. The different coupling strengths, multiple mass scales, 
and multigenerational 
structure should also appear. The Critical Pomeron 
may be the S-Matrix manifestation of the underlying fixed-point field theory. 

If QUD underlies the Standard Model as described, the sextet sector should 
produce new, unmistakeable, large 
cross-sections at the LHC, for which the pomeron could be the main diagnostic!

\end{abstract} 

\renewcommand{\thefootnote}{\arabic{footnote}} \end{titlepage}

\mainhead{ 1. INTRODUCTION.}

To suggest there may be a unique, unitary, particle
S-Matrix is very heretical in
the current theoretical climate - given the wide variety of field theories and
string theories studied. Even so, I will argue that unitarity may be
the key to the origin of the Standard Model as a bound-state S-Matrix embedded
(without off-shell amplitudes) in an almost conformal massless field theory. This is a 
radical proposition which 
the LHC will determine to be either crazy heresy or singularly original insight.

To produce a unitary S-Matrix in
an asymptotically free gauge theory, large momentum perturbation theory
has to match with a high-energy, low transverse momentum, ``non-perturbative''   
solution of both s and t-channel multiparticle unitarity that produces 
asymptotically rising cross-sections. This is an extremely strong constraint, the 
significance of which may not be fully appreciated.
As far as is known, the only non-trivial solution of all high-energy unitarity constraints
is the Reggeon Field Theory Critical Pomeron\cite{cri}. After a long  
quest to understand how this solution can be 
produced by gauge theory reggeon interactions, I have concluded
that a unique underlying massless gauge theory is required. Furthermore, special
small $\beta$-function properties of this theory allow a  high-energy
bound-state S-Matrix, 
dominated by fermion anomalies, to be constructed diagrammatically via 
multi-regge theory. Remarkably, it seems that the S-Matrix of the
full Standard Model could emerge as a generalization of the emergence of 
the hadron S-Matrix from QCD.
The asymptotic scaling of the Critical Pomeron is a reflection of
a fixed-point in the underlying field theory. Here are some quotes from the 
opening paragraphs of a forthcoming paper\cite{amtm}.
\begin{center}
\parbox{6in}{\openup-1.15\jot{ \it 
$~~$ ``~In this paper we will discuss a theory, which we refer to as 
QUD$~^{\#}$, that it appears might provide
a complete and self-contained origin for the Standard Model. We will present 
arguments that SU(5) gauge theory with the 
left-handed, massless, fermion representation\cite{kw}
\centerline{$5\oplus15\oplus40\oplus45^*$}
\newline $~$}}
\parbox{6in}{\openup-1.15\jot{ \it has a bound-state high-energy S-Matrix which
contains only the interactions of the Standard Model and also has,  
qualitatively at least, the correct low mass spectrum.
If the states and high-energy amplitudes are produced by chiral anomaly
coupled multi-regge infra-red divergences, as we outline, then
all elements of the Standard Model will be present
in an extraordinarily economic manner ...}}
\parbox{6in}{\openup-1.15\jot{ \it 
$~~~~$ Although much remains to be done to complete
the picture we develop and many of our arguments are speculative it is clear
that an essential, but 
very unconventional, element that is required for the emergence 
of the Standard Model S-Matrix from QUD is that electroweak symmetry breaking is  
associated with a new, high mass, sector of the (QCD) strong interaction\cite{arw05,
wm,bww}. This new strong sector is predicted\cite{arw05} to  
produce large cross-section effects at the LHC - in addition to 
providing a natural explanation for the existence and 
predominance of dark matter, 
the cosmic ray spectrum knee, and other cosmic ray phenomena.''
\newline $~$
\newline $^{\#}$\hbox{\small Quantum 
Uno/Unification/Unitary/Underlying Dynamics $~$ \hspace{1.3in} $~$}}}
\end{center}

\noindent The discovery of QUD at the LHC would have a 
revolutionary effect on the field !!

\noindent {\bf Experimentally,} the new physics
involves large cross-section phenomena very different from 
common expectations for physics ``beyond the Standard Model''. 
\begin{itemize}
\item{The new phenomena include the strong interaction   
production of both electroweak vector bosons and dark matter 
candidate ``neutrons'' composed of color sextet quarks. 
These cross-sections will be enhanced by 
large (sextet quark) color factors. Sextet neutrons will be stable and their
(QCD) self-interaction will also be
very strong, but with the short-range of the electroweak interaction -
consistent with properties, currently, anticipated for dark matter.}
\item{The ILC would be completely wrong - as the next machine.
A higher-energy SSC would be the obvious choice, preferably with  
help from an e-p machine.}
\end{itemize}
{\bf Theoretically, } QUD 
also has very unexpected and unconventional properties.
\begin{itemize}
\item{As a field theory, it is massless and 
almost conformally invariant. An infra-red fixed-point 
keeps the gauge coupling very small ($\alpha_u 
\raisebox{0.5mm}{${\scriptstyle ~<<~}$}$1/50) 
- providing
a potential explanation for small neutrino masses.}
\end{itemize}
At first sight QUD is an ``unparticle''\cite{hg} theory
that, it would be expected\cite{asv}, can not have a non-perturbative particle spectrum 
because of the infra-red scale invariance of off-shell Green's functions.
Our expectation is, however, that the full field theory is defined only perturbatively 
(as a large momentum expansion), with no well-defined non-perturbative 
Green's functions, and in a major break with the current theoretical 
paradigm, we expect the physical states and interactions to appear only in the S-Matrix.  
Infra-red scale invariance is then manifest in the chiral anomaly coupled
wee gluon reggeon interactions that dominate the dynamics 
producing a multi-regge bound-state S-Matrix. Amongst the significant properties 
that emerge are  
\begin{enumerate}
\item{Anomaly domination of wee gluon reggeon interactions implies that
only a very small subset of the field theory degrees of freedom contribute to the
S-Matrix.}
\item{An ``anomalous wee gluon'' critical phenomenon occurs in which,
because of the conjugacy properties of the fermions, only Standard Model 
interactions survive.} 
\item{All particles are, Goldstone boson related, bound-states with
masses generated by reggeization, mixing, and anomaly interactions  
{\bf - there is no Higgs field.}}
\end{enumerate}
 
It is well-known that, at infinite momentum, ``universal wee partons'' could, potentially,
play the role of a vacuum and that this is probably a necessity for a full parton
model to be valid in QCD. In fact, we are able to access
``non-perturbative'' physics in the multi-regge
region just because, in effect, infinite momentum frame kinematics are  
introduced for all states and interactions. (As a result, our description of the 
properties of states can be quite different from, although it must be consistent
with, descriptions at finite momentum.) By using the power of the multi-regge theory
that we have developed, we are then able to construct bound-state amplitudes in terms
of reggeon diagrams. Previously we have shown\cite{arw05} that in  
massless QCD$_S$, a QCD sub-sector of QUD, the chiral anomaly dynamics produces
an anomalous wee gluon ``vacuum'' component of 
both the pomeron (which is critical) and all bound-states. Although there are 
important distinctive properties relative to conventional QCD, as far as we know,
our results are consistent with all the (experimentally established) 
properties of QCD below the electroweak scale.
A crucial distinction is, however,
the limitation on the spectrum of states compared
to what would be anticipated from just color confinement and chiral
symmetry breaking. The spectrum we obtain, particularly
the absence of glueballs , is more consistent with experiment than  
conventional QCD expectations, as is the absence of the BFKL pomeron. 

In the following, we will outline arguments that the chiral anomaly dynamics 
responsible for our solution of QCD$_S$, produces in QUD just the 
interactions and bound-states of the Standard Model. As will become apparent, 
unfortunately, although we are able to describe qualitatively how the dynamics 
operates and how states and amplitudes are obtained, there is still much that needs to
be done to indentify physical scales and even to determine how many parameters are
actually involved in the constructions we describe. 

We build up the high-energy behavior of both QCD$_S$ and QUD by starting in a 
supercritical pomeron (color superconducting) phase in
which the gauge symmetry is partially broken. A resulting 
anomalous gluon infra-red divergence produces a wee gluon condensate
and directly determines
the physical amplitudes, with the physical states being anomaly-pole chiral 
Goldstone bosons. Higher-order contributions of chirality 
transition vertices can then be built up perturbatively. 
Restoration of the full color symmetry produces critical behavior
in which the production and absorption of anomalous wee gluons becomes a dynamical 
 collective phenomenon.  

The vertices that couple anomalous wee gluons contain  
both a chirality transition and an on-shell longitudinal gluon interaction.
This interaction is present in the color superconducting phase and can be present
in the unbroken theory via the Gribov (light-cone) quantization ambiguity.
Physically, the anomaly vertices 
describe the production of a fermion pair, one of which is a zero
momentum hole state which becomes physical
via compensating wee gluon emission. 
Viewing the wee gluon emission as resulting from an initial
displacement of the Dirac sea, we can identify 
this ``Fermi surface fluctuation'' as the order parameter of the pomeron phase-transition. 
In the supercritical phase it is  a correlated 
``reggeon condensate'' that, in our construction, is 
introduced by the color symmetry breaking. 
At the critical point the Fermi surface fluctuations
are dynamical and uncorrelated (locally random within the color group), but produce
long-range correlations as a collective phenomenon. In QUD,
the Fermi surface anomalous wee gluon emission (and Critical Pomeron behaviour) 
is limited to a non-abelian subgroup with vector-like 
couplings to the sea. Because QUD
is vector-like with respect to an SU(3)xU(1) subgroup only, 
the parity conserving SU(3) color strong interaction emerges. 

We will see that QUD is an extraordinarily minimal extension of the Standard 
Model in that almost
all of the elements that are clearly ``Beyond the Standard Model'' have an essential
dynamical role. There is 
a sextet quark sector that, as we have emphasized, produces 
both electroweak symmetry breaking and dark matter. There is also 
an octet quark sector that has lepton electroweak quantum numbers. As 
a real SU(3) representation, the octet quarks play a crucial role
in allowing leptons to be SU(5) invariant, while having no strong interaction. 
The octet sector also appears to be 
responsible for the emergence of the Standard Model generation structure.

\mainhead{2. COSMIC RAY AND TEVATRON EVIDENCE} 

There is already substantial evidence in cosmic ray physics that there could be a
major strong interaction change in the energy range between the Tevatron
and the LHC. In Fig.~1 we show the well-known knee in the 
cosmic ray spectrum.
It is a remarkable, very well-established, phenomenon that occurs between
Tevatron and LHC energies. The associated break in 
the slope stands out, distinctively, as the 
energy increases over some ten orders of magnitude and the flux decreases
by thirty orders of magnitude. 
Although it is generally believed to be due to 
a (so far not understood) conspiracy of
external phenomena, a change in the atmosheric interaction seems far more plausible. 
From it's earliest discovery, it was 
suggested\cite{nik} that the knee could be a threshold for
production of 
neutral particles not observed in the ground level detectors.
The resulting underestimation of the shower energy 
would lead to a pile-up of events that would be
observed as a ``knee''. However,
a major part of the cross-section has to be involved and there was no credible
proposal as to what the neutral particles could be.

The energy scale for the sextet sector is the electroweak scale and 
sextet cross-sections should be larger than triplet cross-sections because of color
factors. However,
the effective energy threshold for production of the sextet sector is determined by
inclusive pomeron exchange, since this is the only large cross-section mechanism for
sextet states to be produced by triplet states. According to current
diffractive phenomenology, the onset of inclusive pomeron production of
electroweak scale states should be around
the energy of the knee. Three effects that will contribute to the formation of a knee
are 
\newline $~$
\newline 
\parbox{3.1in}{\epsfxsize=3in
\epsffile{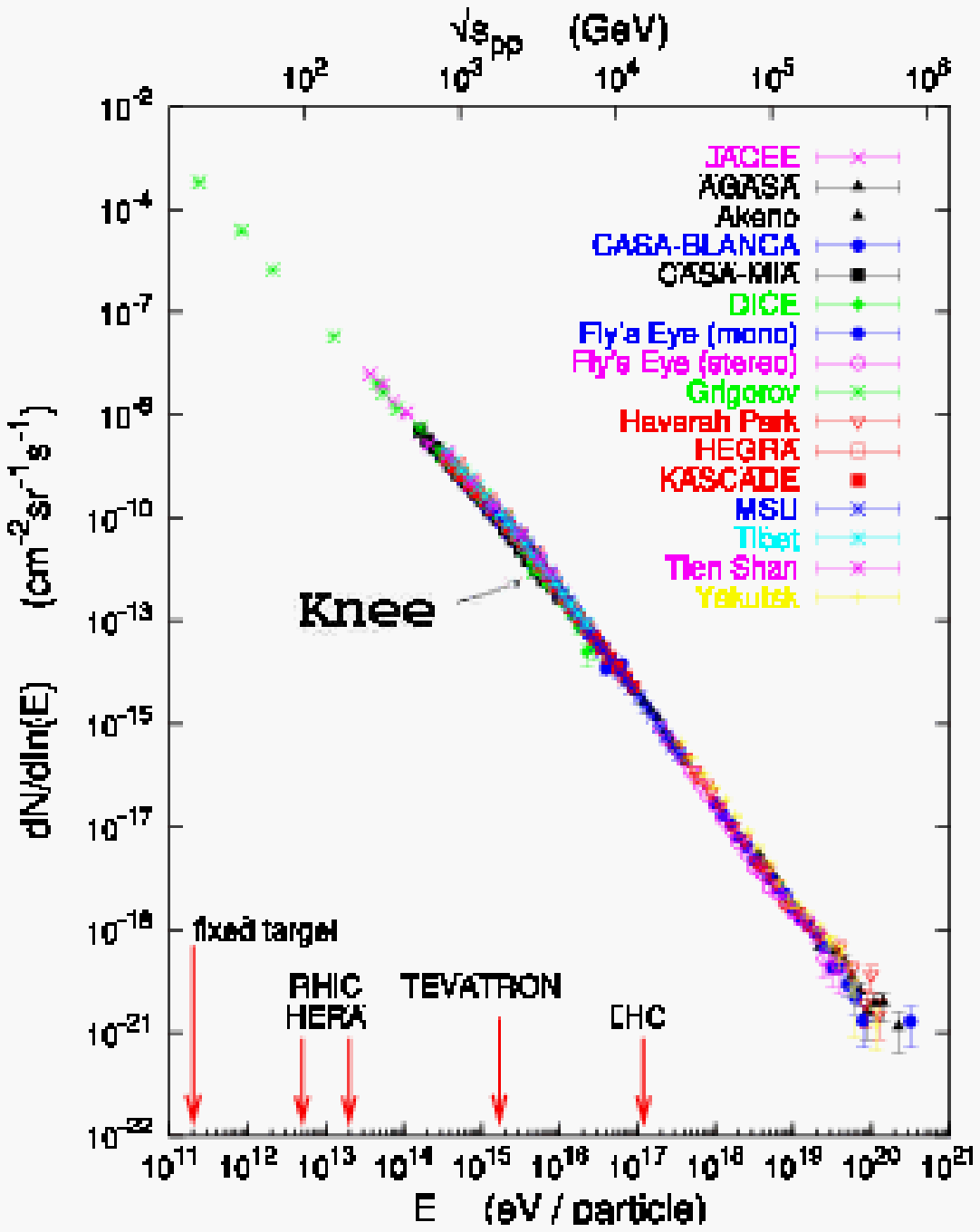}
\centerline{(a)}}
\hspace{0.1in}
\parbox{2.3in}{
\epsfxsize=2.2in
\epsffile{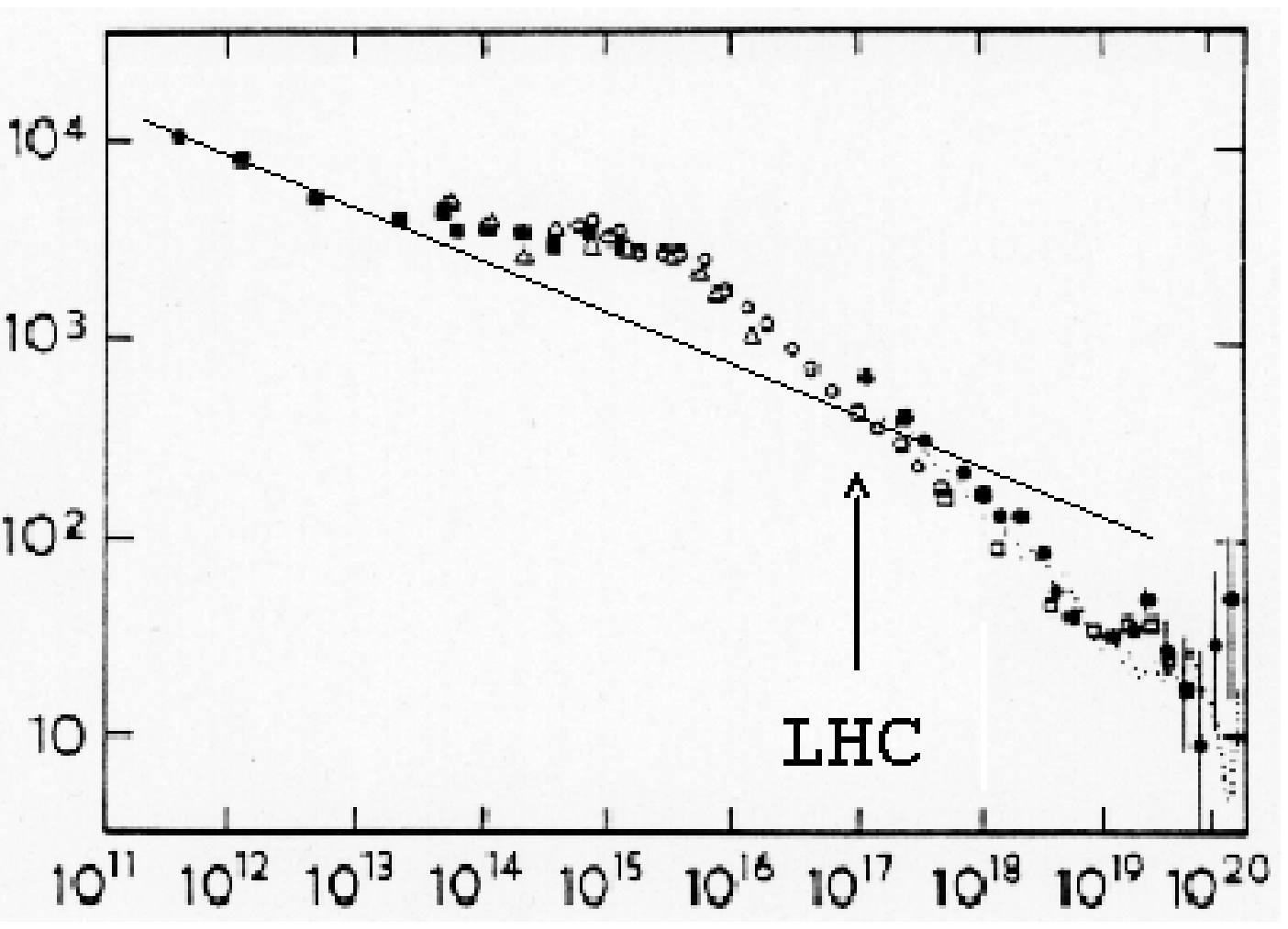}
\centerline{(b)}
\epsfxsize=2.2in
\epsffile{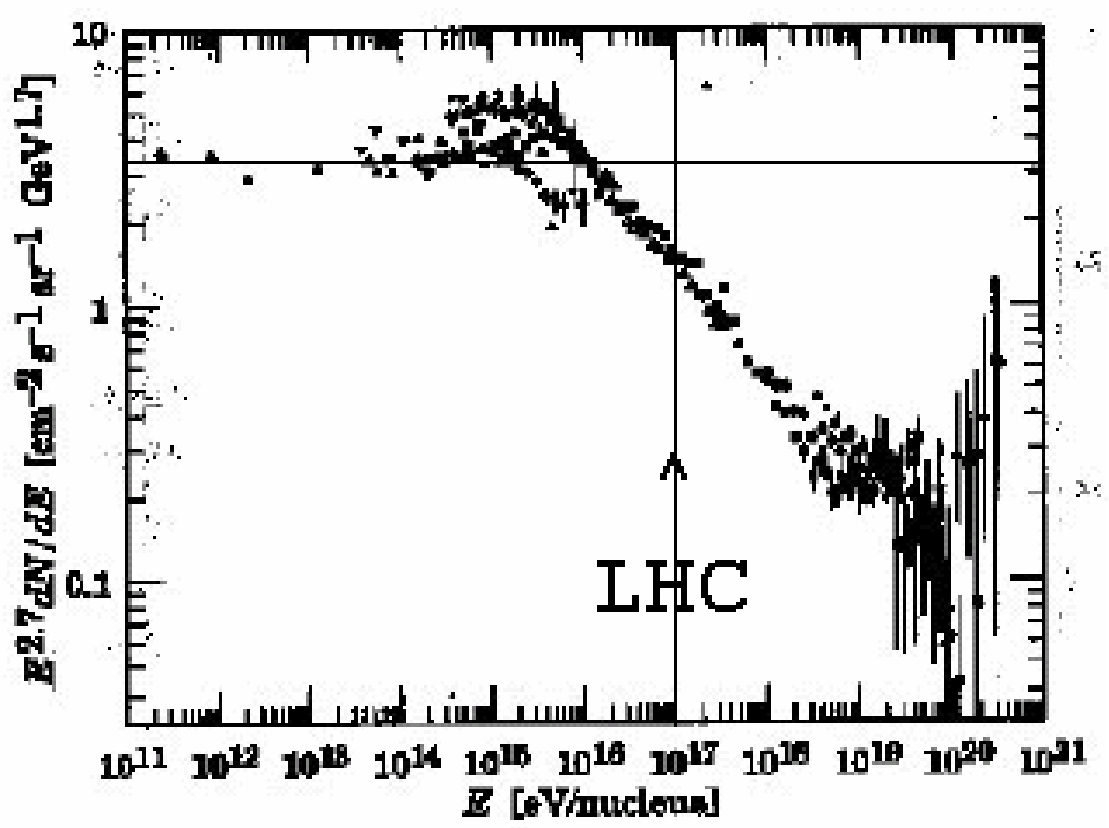}
\centerline{(c)}}
\begin{center}
Figure 1. 
The knee (a) all data (b) less data (c) with the slope extracted. 
\end{center}
\begin{enumerate}
\item{Prolific production of vector bosons 
will increase the average transverse momentum enormously and will also
increase the relative neutrino production. Energies will
be seriously underestimated by the unexpected shower spread and increased neutral
component. } 
\item{Dark matter (sextet neutron) production, and the resultant energy underestimation,
will increase rapidly with
energy. This is close to the original explanation 
of the knee as due to the production of neutrals. Of course,
dark matter was unknown and the link with the knee, that I am proposing, 
could not have been imagined.}
\item{Sextet neutron dark matter should be a major component of incoming cosmic rays. Since
the pomeron again has to be involved, the atmospheric interaction will have
a threshold energy that is
not far below the normal matter sextet threshold energy and, once underway, will share
properties 1. and 2.}
\end{enumerate}
Probably, the observed knee can only be reproduced if
this last effect (which will produce a direct ``bump'' in the spectrum) 
is a significant part of it's formation.
Unfortunately, we have ignored this effect 
in our previous discussions. Consequently, using the knee events as a basis, our estimates
of the needed magnitudes for the first two effects, and the corresponding LHC 
cross-sections, were dramatically large and, perhaps,
impossible to reproduce theoretically. With the third effect included, the knee seems
relatively straightforward to reproduce. The expected  
new LHC cross-sections must still be large, strong
interaction, effects. However, they may not be quite as ``dramatic'' as we have 
previously emphasized.

There are a large number of other new phenomena seen in cosmic ray showers
with energies above the knee. In Fig.~2 we show one of the most interesting\cite{cores},
involving  the production (essentially) of high $E_T$ jet pairs. 
A QCD Monte Carlo 
tuned to jet data at collider energies fails to reproduce the data 
above the knee, by orders of magnitude, as would be expected if the sextet sector 
is produced.
\begin{center}
\epsfxsize=2.9in
\epsffile{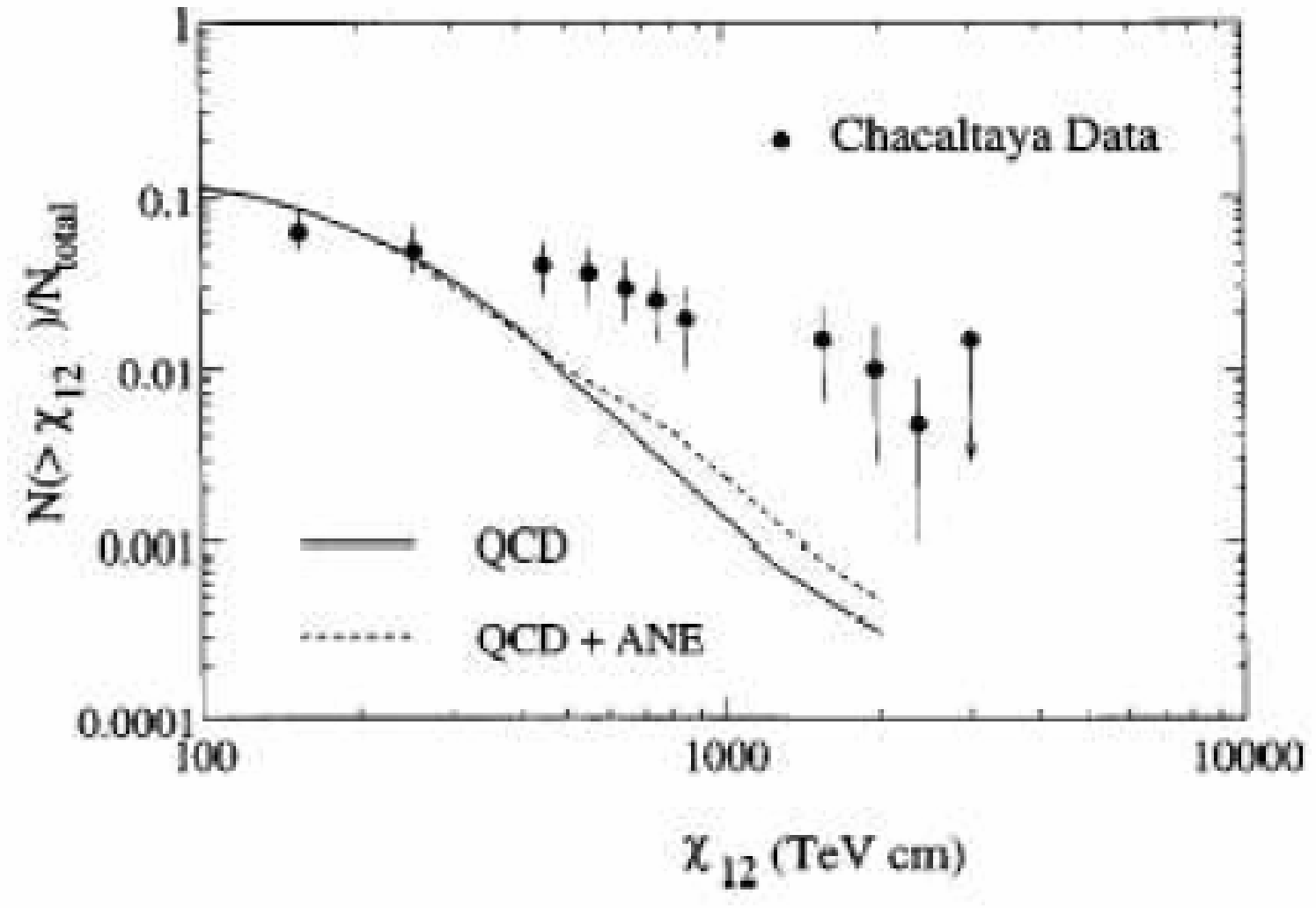}

Figure 2 Dijet production
\end{center}
There are also indications from the 
Tevatron that there will be a strong interaction change at higher energies. 
We discuss two phenomena,
illustrated in Fig.~3, 
\begin{center}

\epsfxsize=3in
\epsffile{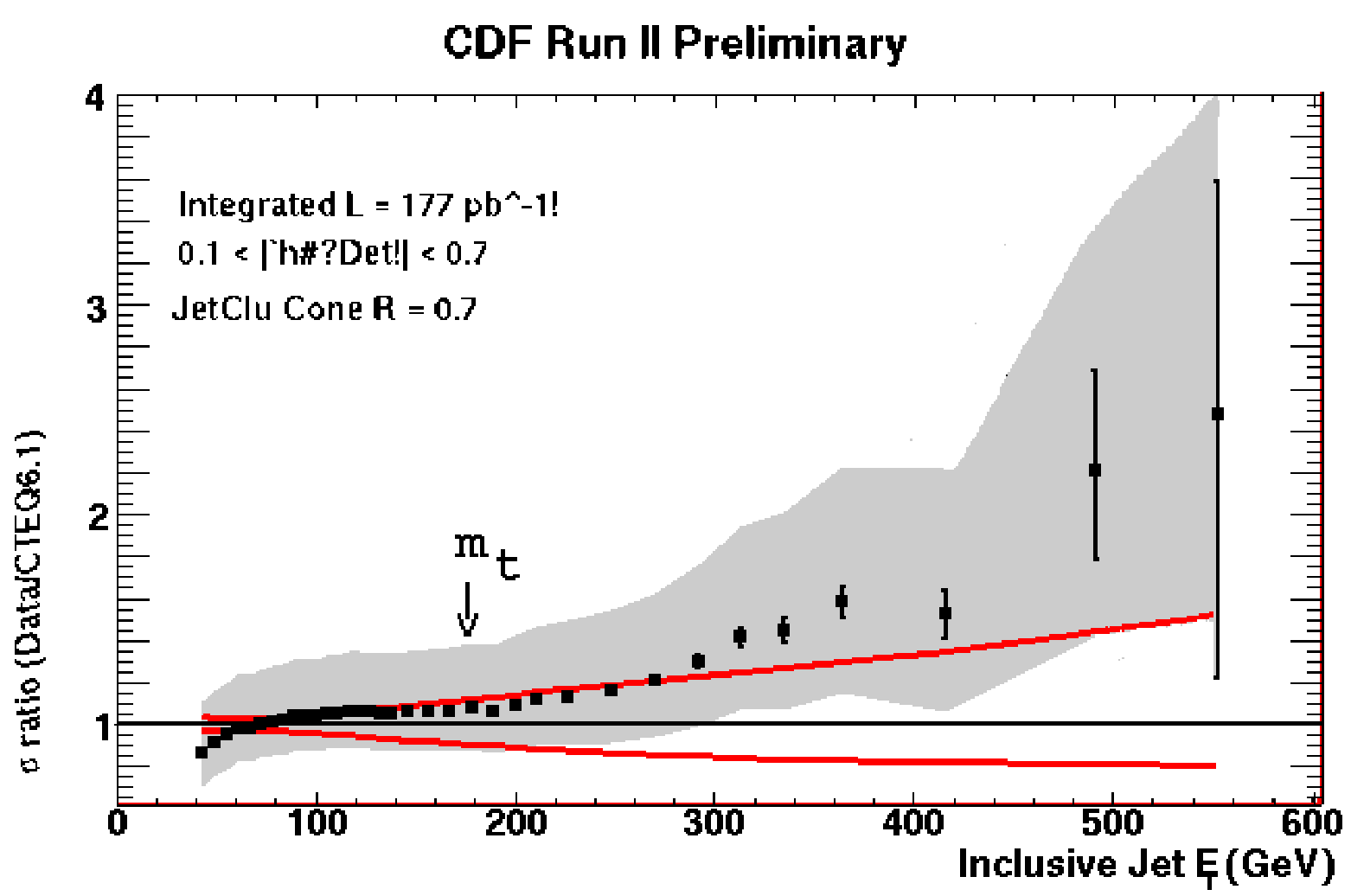}
\hspace{0.3in}
\epsfxsize=2.4in
\epsffile{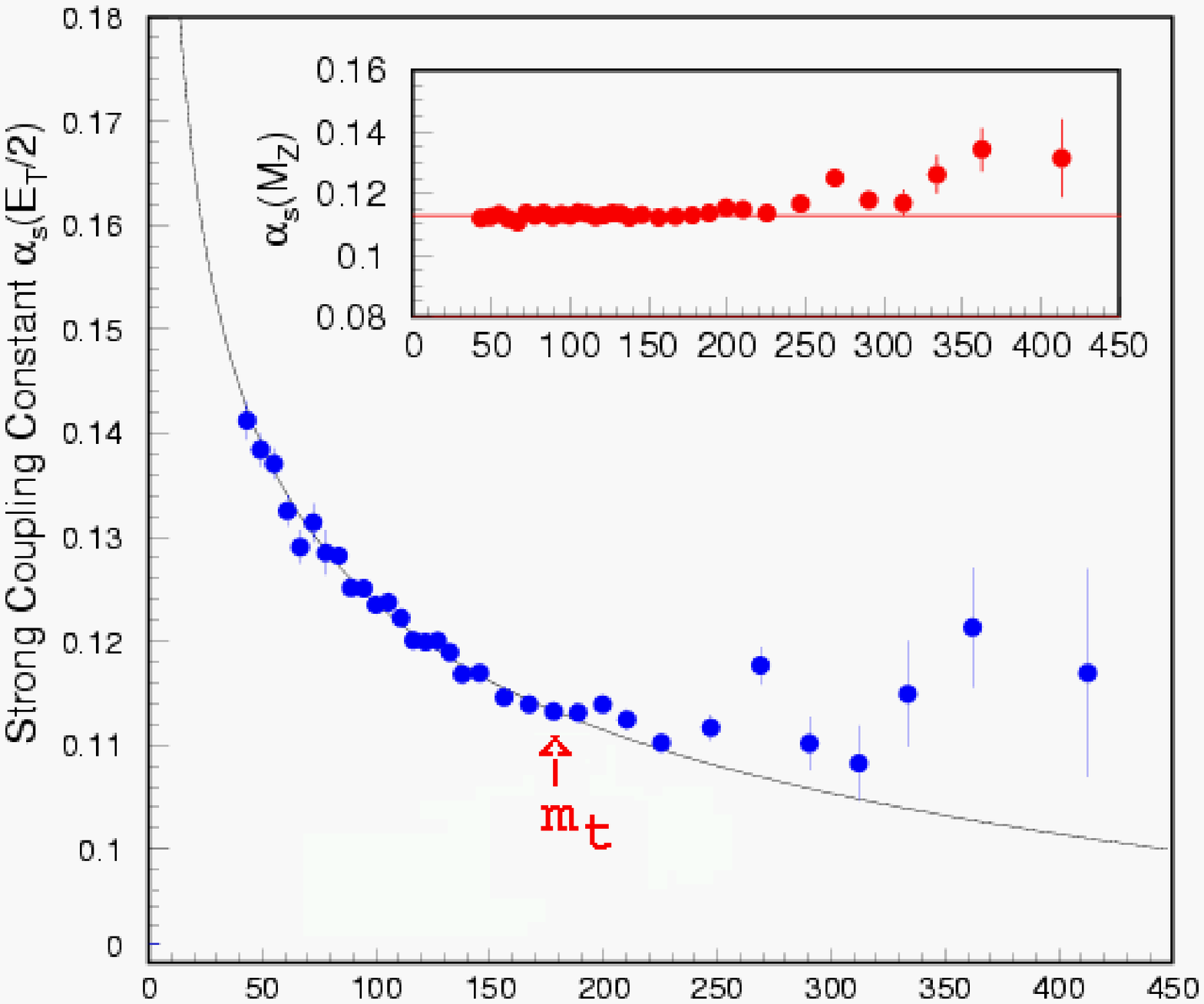}

(a) \hspace{3in} (b)

Figure 3(a) The Inclusive Jet Cross-Section (b) Evolution of $\alpha_s$

\end{center}
which both suggest that new QCD physics enters above
the top mass scale.
In Fig.~3(a) we show the Run 2 inclusive jet cross-section\cite{CDFb} 
obtained using a cone algorithm. Naively, the data 
pull away from the theory from $E_T \sim m_t$ upwards, indicating that
QCD jet physics above the electroweak scale
may be breaking down in just the manner
that we would expect, as the sextet sector enters the theory.

In Fig.~3(b) we show the measured evolution\cite{CDF} of $\alpha_s$. The inclusion of 
a sextet quark doublet in the 
QCD $\beta$-function would halt the evolution of $\alpha_s$, 
just as appears to be happening
at $E_T \sim m_t$. That this is the right sextet scale can be argued by 
noting that $E_T \sim m_t$ is also $E_T \sim 2 M_W$. Alternatively,
as discussed at more length in \cite{arw05}, it could
well be that top production is due to a resonance (the $\eta_6$) that is the sextet
sector analog of the Higgs particle. This would imply that the top quark mass is actually
the sextet constituent mass scale, and it would be expected that deviations from
conventional QCD would start above this scale.

\mainhead{3. THE PATH TO CONFINING QUD}

The logical steps that lead, uniquely, to QUD can be summarized 
as follows. They will 
be discussed in much more detail in \cite{amtm}.
\begin{enumerate}  
\item{The Critical Pomeron, obtained\cite{cri} as a renormalization group solution of
Reggeon Field Theory (RFT), is the only known 
solution\cite{mm} 
of full multiparticle unitarity - in both the $t$ and $s$ channels - that gives
asymptotically rising cross-sections.}
\item{Supercritical RFT matches\cite{arw05} 
with color superconducting QCD. That hadrons must also have 
supercritical properties\cite{arw05} is achieved, as we will see, by 
``anomalous wee gluons'' appearing in both hadrons and the pomeron. The matching
with supercritical RFT, shows that the Critical Pomeron occurs in QCD
when asymptotic freedom is ``saturated''. 
The only realistic quark content is six color triplets
plus two color sextets, giving\cite{arw84} what we refer to as ``QCD$_S$''}
\item{If the sextet quarks have the right electroweak quantum numbers,
the ``sextet pion'' sector will produce electroweak symmetry breaking 
- without any new interaction. The electroweak scale will be the QCD
sextet chiral scale, which Casimir scaling implies\cite{arw05,wm} 
is the right order of magnitude.}
\item{To cancel the electroweak
anomaly and to generate particle masses, the sextet sector should 
be embedded in a left-handed unified theory.}
\item{Asking for the sextet sector, plus asymptotic freedom, plus anomaly cancelation,
uniquely\cite{kw} selects QUD, i.e. SU(5) gauge theory 
with the left-handed fermion representation 
\newline \centerline{\bf $~~5\oplus15\oplus40\oplus45^*~~~~~~~~~~$}}
\end{enumerate}

Amazingly, the triplet quark and lepton sectors of QUD,
although they were not asked for, are very close to the Standard Model !!
There are three ``generations'' of 
quarks and antiquarks with charges $~\pm 2/3,~ \pm 1/3~$, (implying that
QUD contains QCD$_S$) and there are also three ``generations'' of leptons.
The $SU(3)\otimes SU(2)\otimes U(1)~$ decomposition of QUD is
{\small 
$$
\eqalign{5~&=(3,1,-\frac{1}{3}))^{ \{3\}}
+(1,2,\frac{1}{2}))^{ \{2\}}~,\cr
15~&=(1,3,1)~+~(3,2,\frac{1}{6})^{ \{1\}}+
{\bf (6,1,-\frac{2}{3})}~,\cr
40~&=(1,2,-\frac{3}{2})^{ \{3\}}
+(3,2,\frac{1}{6})^{ \{2\}}+
(3^*,1,-\frac{2}{3})+(3^*,3,-\frac{2}{3})+
{\bf (6^*,2,\frac{1}{6})}+(8,1,1)~,\cr
 45^*&=(1,2,-\frac{1}{2})^{ \{1\}}+(3^*,1,\frac{1}{3})
+(3^*,3,\frac{1}{3})+(3,1,-\frac{4}{3})
+(3,2,\frac{7}{6})^{ \{3\}}+
{\bf (6,1,\frac{1}{3})} +(8,2,-\frac{1}{2})}
$$}
The ``Standard Model'' quark and lepton generations, denoted by superscripts  
$~\{1\},~\{2\}$, and $\{3\}$,
are scattered amongst the separate SU(5) representations.
Clearly, the SU(2)xU(1) quantum numbers are not quite right 
when compared directly with the Standard Model. For a long time this seemed to be
an insuperable problem for relating QUD to the Standard Model via any standard
(or non-standard) Higgs' mechanism.
Note that QUD is real, i.e. is a vector theory,
with respect to SU(3)xU(1)$_{em}$.

Eventually, I realized that QUD should be considered directly as a 
confining theory, without any additional symmetry breaking mechanism,
and that this is how it can compare 
with the Standard Model. In the QUD S-Matrix SU(5) color is confined, 
not just SU(3) color, and so 
all elementary gauge bosons and fermions are confined and massless.
For the Standard Model to emerge, it must have
the same relationship to QUD that the hadronic sector has to QCD !!!
All hadrons and leptons have to be QUD bound-states and, also, 
all Standard Model interactions have to be composite. 

That confining QUD might give the Standard Model 
S-Matrix became apparent to me only after I understood\cite{arw05} 
the dynamics of high-energy massless QCD$_S$ in which, as we have
described in the Introduction, the S-Matrix is dominated by 
anomaly vertices containing zero momentum fermion chirality transitions.
The chirality transitions play a similar role to condensates - but only in the S-Matrix~!!
In our multi-regge construction of amplitudes they appear
as relic effects in anomalous reggeon vertices when initial mass generating
scalar fields are decoupled. (The description of the 
dynamics that we give will make it clear why the 
chirality transitions contribute only in scattering processes involving  
asymptotic states.)
In massless QCD$_S$, because it is a vector theory that conserves parity, the chirality 
transitions lead only to chiral 
symmetry breaking and color parity breaking by the pomeron.
Correspondingly, it is because QUD is vector-like only with respect to 
an SU(3)xU(1) subgroup
that the chirality transition anomaly vertices naturally select the interactions 
of the Standard Model.

The only elements of the QUD fermion representation 
that are clearly ``Beyond the Standard Model'' are the following.
\begin{enumerate}
\item{The sextet quark sector, as we have already
talked about, produces both electroweak symmetry breaking and dark matter.}
\item{The octet quark sector has lepton electroweak quantum numbers. Because they carry
a real SU(3) representation, the octets 
can not form physical states as chiral Goldston bosons.
As a result, they play a very different dynamical role to the triplet and sextet quarks.
As we will see, they are nevertheless crucial
in allowing leptons to be SU(5) invariant and in 
producing the correct generation structure of the Standard Model.}
\item{We will not find a role for a 
pair of exotically charged triplet quarks.}
\end{enumerate}

\mainhead{4. MULTI-REGGE REGGEON AMPLITUDES}

The analytic multi-regge theory that we have developed over the years\cite{arw98}
provides a very powerful tool for the construction of
bound-state scattering amplitudes. 
For example, the amplitude for bound-state regge pole pions to scatter via pomeron 
exchange should be contained in a di-triple-regge amplitude, as illustrated in Fig.~4.
\begin{center}
\epsfxsize=2.5in
\epsffile{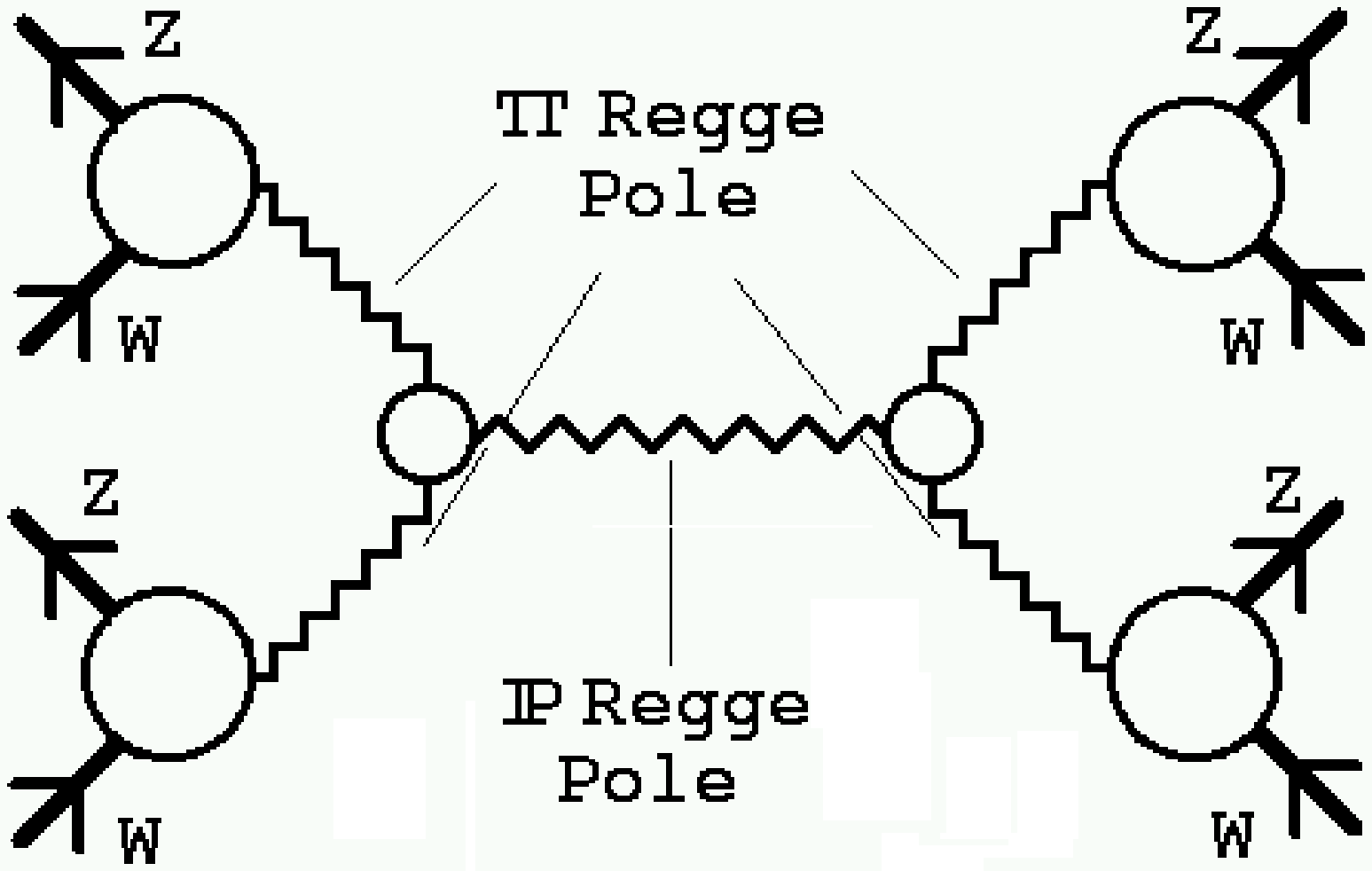}

Figure 4. A di-triple-regge amplitude.
\end{center}
(It will be clear later why we have chosen the 
external scattering states to be vector bosons.)
Kinematic descriptions of multi-regge limits can be found in various 
references\cite{arw98} and we will not give them here. Suffice it to say that
each circle vertex in Fig.~4 is separated from adjacent vertices
by a large longitudinal momentum, in various space directions, in such a way that
a regge exchange contributes to each leg of the diagram. (As we noted in the Introduction,
this implies that both bound states and interactions can be viewed as infinite
momentum states that, as we will see, have a ``wee parton'' component that 
has a vacuum-like role.) Here 
we will concentrate on the procedure for obtaining an 
amplitude of the form of Fig.~4 
from QCD$_S$ and QUD reggeon diagrams. 
What we describe will be just an outline. More details will be given in \cite{amtm}.
We anticipate that, if QUD is discovered, the use of reggeon diagrams to calculate
bound-state amplitudes will become a major technology and what we describe would be just
the beginning of what would become the central calculational procedure.

The multi-regge region is where the abstract 
properties of the S-Matrix are the most powerful\cite{arw98,arw00}.
In this kinematic regime, S-Matrix amplitudes
have a simple analytic structure and the physical region
discontinuity formulae needed to write multiple dispersion relations have been
established. As a consequence, multiparticle complex angular momentum theory
has been put on a firm foundation and,  
most importantly, the multi-regge S-Matrix has been shown\cite{arw98} to be 
controled by ``reggeon unitarity'' equations 
formulated directly in the angular momentum plane. These equations are 
discontinuity formulae for all the singularities that control multi-regge asymptotic
behavior. Before QCD, these equations were used to formulate Reggeon Field Theory
as an effective lagrangian formalism which perturbatively solves the unitarity
equations. A renormalization group formalism was then introduced and the Critical
Pomeron obtained\cite{cri} as a fixed-point solution.  
That the reggeon unitarity equations are satisfied by the abstract, but 
calculable, RFT Critical Pomeron can be regarded as
the pinnacle achievement, so far at least, of abstract S-Matrix Theory.

Reggeon unitarity is also satisfied, perturbatively, by all existing calculations
of gauge theory regge behavior.
The leading multi-regge behavior of a feynman diagram is typically obtained
by routing the large external light-cone momenta 
through the diagram so that the 
maximal number of particles are close to mass-shell and have 
large, relative, longitudinal momentum separations. 
After longitudinal integrations are carried out, the result is a transverse momentum 
diagram (integral) multiplied by logarithms of invariant energies.
In a non-abelian gauge theory, all the transverse 
momentum diagrams generated perturbatively can be organized\cite{bs} into 
gluon and quark reggeon diagrams.
Reggeon diagrams are transverse momentum diagrams 
with additional reggeon propagators (in either complex angular momentum 
or rapidity space) that 
reproduce the energy logarithms while simultaneously satisfying reggeon unitarity.  

As far as is known, when all the reggeons are massive, reggeon diagrams 
provide a complete, perturbative, description of a
spontaneously broken gauge theory in all multi-regge limits. In particular, 
this is the case 
when the gauge symmetry is completely broken by the Higgs mechanism.
The reggeon diagrams are gauge 
invariant but both the reggeized gauge bosons and 
the reggeized fermions carry their original gauge theory representations as global 
representations of the gauge group.
In general, Higgs scalars do not reggeize and so (before the Higgs scalars
are decoupled) multi-regge theory is applicable only to the leading high-energy 
behavior which originates from the vector bosons and fermions. 

To obtain the bound-state amplitude of Fig.~4 it is necessary, a priori, to
consider all multi-regge reggeon diagrams of the general form illustrated 
in Fig.~5, and a lot more as well. At first sight, this an impossibly 
difficult task.
\begin{center}
\epsfxsize=4.5in
\epsffile{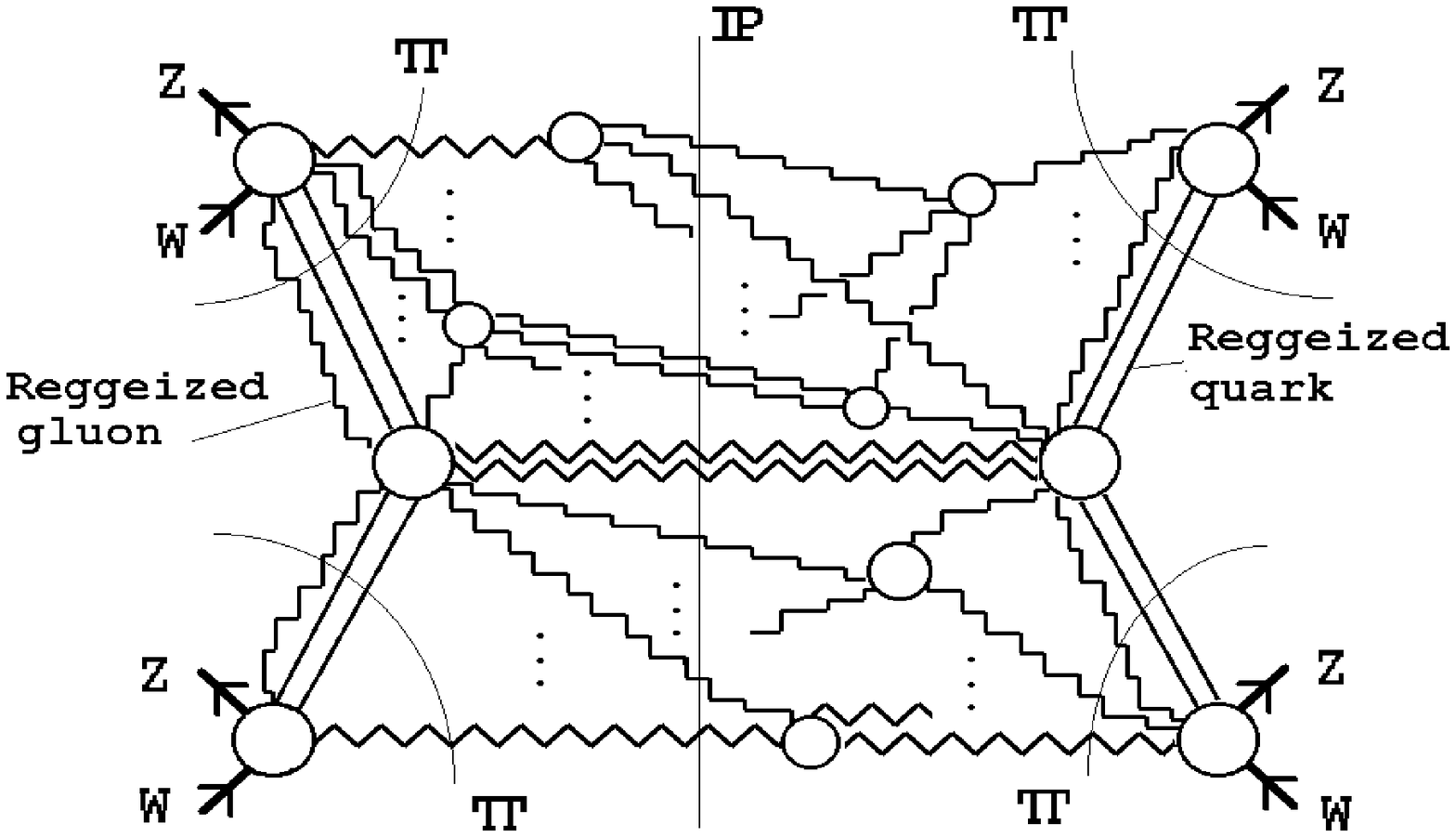}

Figure 5. A di-triple-regge reggeon diagram amplitude.
\end{center}
 Fortunately, the problem
can be brought under control as follows. Reggeon unitarity 
determines\cite{arw98} that 
the kinematic structure and interactions in each t-channel
are the same as in well-known elastic scattering diagrams. Only 
the interactions coupling the separate t-channels (the largest circles in Fig.~5)
are more complicated. As a result, we can discuss
the infra-red divergences that occur when the reggeons are massless in general terms before
considering the specific, very complicated, diagrams. We can then show that the
divergences leave only a much smaller set of diagrams, with very special properties.
Most importantly, in all the surviving diagrams, the largest circle amplitudes
in Fig.~5 contain the anomaly 
vertices that we alluded to in the Introduction and describe further 
in the next Section.

To obtain our starting point of massive gauge boson
and fermion reggeons we use fundamental representation scalar fields. 
According to complimentarity\cite{fs}, this
ensures a smooth restoration of the underlying gauge symmetry when there is
a $k_{\perp}$ cut-off in place. 
(Note that it is the properties of reggeon diagrams in 
the $k_{\perp}$ infra-red region that are important for our purposes,
in contrast to the large $k_{\perp}$ significance\cite{fl} of such diagrams
in BFKL physics.) The problem then is   
to understand how, and in what circumstances,  
can the massive reggeon diagram solution of
reggeon unitarity, that perturbatively describes 
a spontaneously broken gauge theory, convert to a bound-state and pomeron diagram
description containing amplitudes of the form of Fig.~4,
as the gauge symmetry is restored. Most importantly, of course, 
we want to understand the circumstances that will give the Critical Pomeron.
We expect that the emergence of Critical Behavior will be very important in allowing 
the decoupling mass and cut-off scales, that we start with, to be replaced by the
scales of the massless theory.

\mainhead{5. INFRA-RED DIVERGENCES AND ANOMALIES}

To give a general idea of what all the contributing factors are, we will keep 
the following description of our infra-red divergence analysis very qualitative.
The technical details will be described in much more detail in \cite{amtm}.

It is well-known that the massless limit produces exponentiating infra-red 
divergences, associated with reggeization, that 
``confine'' the global color in the sense that only 
reggeon diagrams containing color zero combinations of reggeons survive.
However, this is not real confinement in that color zero multi-gluon singularities remain.
In our procedure we carry out the ``confinement'' of (global) color in stages. 
We impose a $k_{\perp}$ cut-off until the last stage when, as we
discuss in the following Section, 
an asymptotically-free scalar field can be used. A major consequence of
the cut-off is that fermion loop reggeon interactions
do not have Ward identity zeroes at $k_{\perp} =0$ and, as a result,
additional infra-red divergences appear. Almost all of the additional divergences
exponentiate and so, because of the cut-off, 
a much larger part of the theory, beyond just the color non-zero
sectors, is removed. As we will see, there is a dominant surviving divergence,
due to the fermion anomaly vertices that we discuss next,
that does not exponentiate. Because this divergence does produce genuine
confinement we will be able to simply refer to a massless limit 
that restores a global color symmetry as confining that symmetry. 
 
Reggeon interaction vertices that contain anomalies couple 
``anomalous gluons''. As we will
define explicitly in Section 7, ``anomalous gluons'
are combinations of gluon reggeons that are
reggeon generalizations of the well-known anomaly current.
The anomalies appear in effective triangle diagrams that 
are produced when fermions in large loops are placed on-shell by a multi-regge limit.
Because multi-regge limits are defined for on mass-shell amplitudes, the 
anomalies that are generated are a strictly S-Matrix phenomenon.
Examples of anomaly vertices (derived in my papers) are shown in Fig.~6.

The first vertex shown appears in the coupling of a scattering vector boson
to the combination of fermions and anomalous gluons that becomes 
a regge pole pion as we describe below. As is also illustrated, in this vertex
an (on-shell) longitudinal vector interaction plays an essential
role in producing the effective triangle diagram. This is one major reason
why starting with 
massive reggeons is a necessary part of producing the anomaly vertices.
Because anomalies require three-dimensional kinematics they 
can only occur in vertices coupling external states or in vertices coupling
distinct reggeon t-channels. Therefore, they can appear in the vertices 
represented by circles in Fig.~4 and by the largest circles in Fig.~5, but they 
can not occur within the self-interactions of a multi-reggeon configuration 
forming a single t-channel reggeon state (i.e. the pion and pomeron in Figs.~4 and 5). 

A further property of an
anomaly vertex, that has a deep significance for our construction of bound-state
amplitudes, is also illustrated in the first vertex of Fig.~6.
An ``anomaly pole'' is generated\cite{arw02} 
when the anomalous gluons involved carry $k_{\perp} = 0$. 
As illustrated, production
of the pole involves a zero momentum chirality transition, that would be the  
zero-momentum contribution of a propagator 
to a condensate. The presence of such chirality
transitions is another consequence of starting with
masses for all fermions and gauge bosons that is also 
crucial for the generation of anomalies. 
The chirality
transitions are effects of the initial masses which remain in the anomaly vertices 
after the masses are sent to zero. We emphasize that the masses decouple 
straightforwardly in all non-anomaly reggeon interactions and therefore
the associated large $k_{\perp}$ perturbation theory is given by the massless theory. 

\begin{center}
\epsfxsize=5.8in
\epsffile{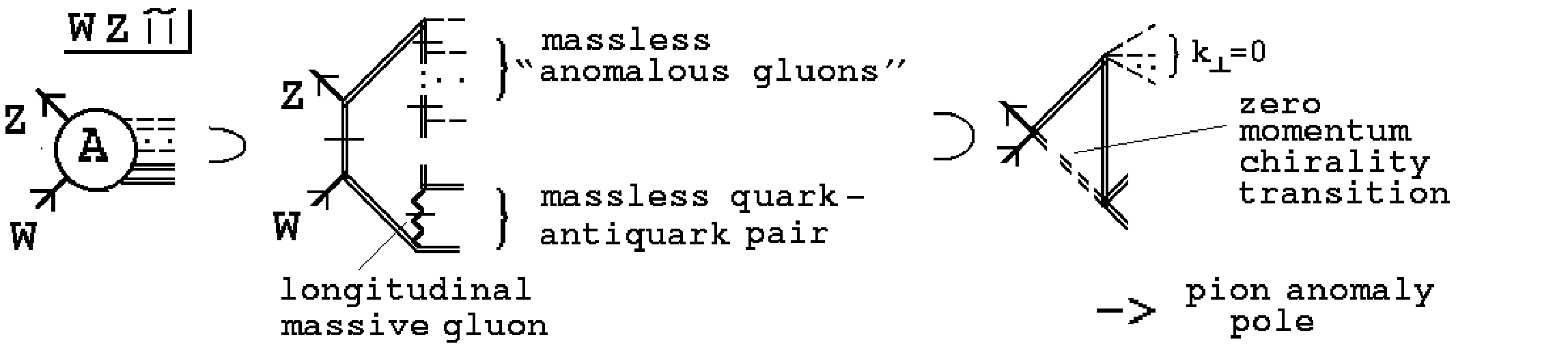} 
\newline$~$
\newline \epsfxsize=5.5in
\epsffile{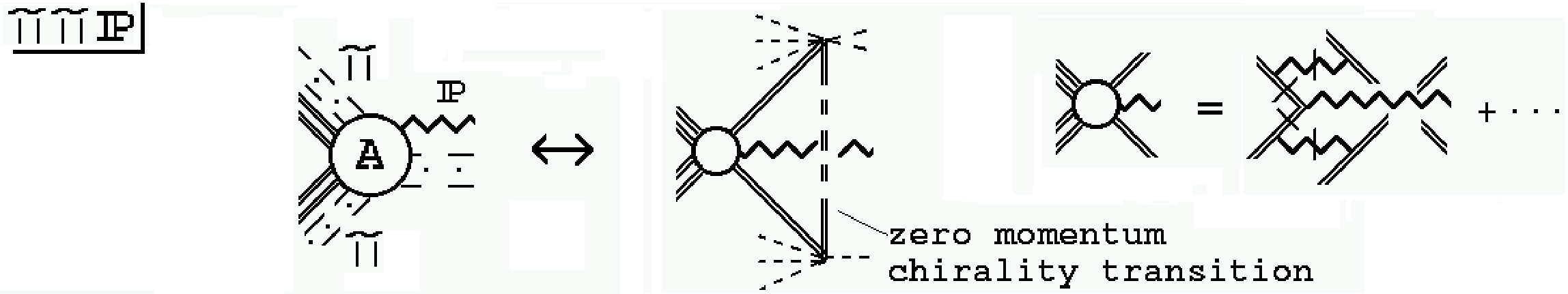}
\newline $~$
\newline \epsfxsize=1.6in
\epsffile{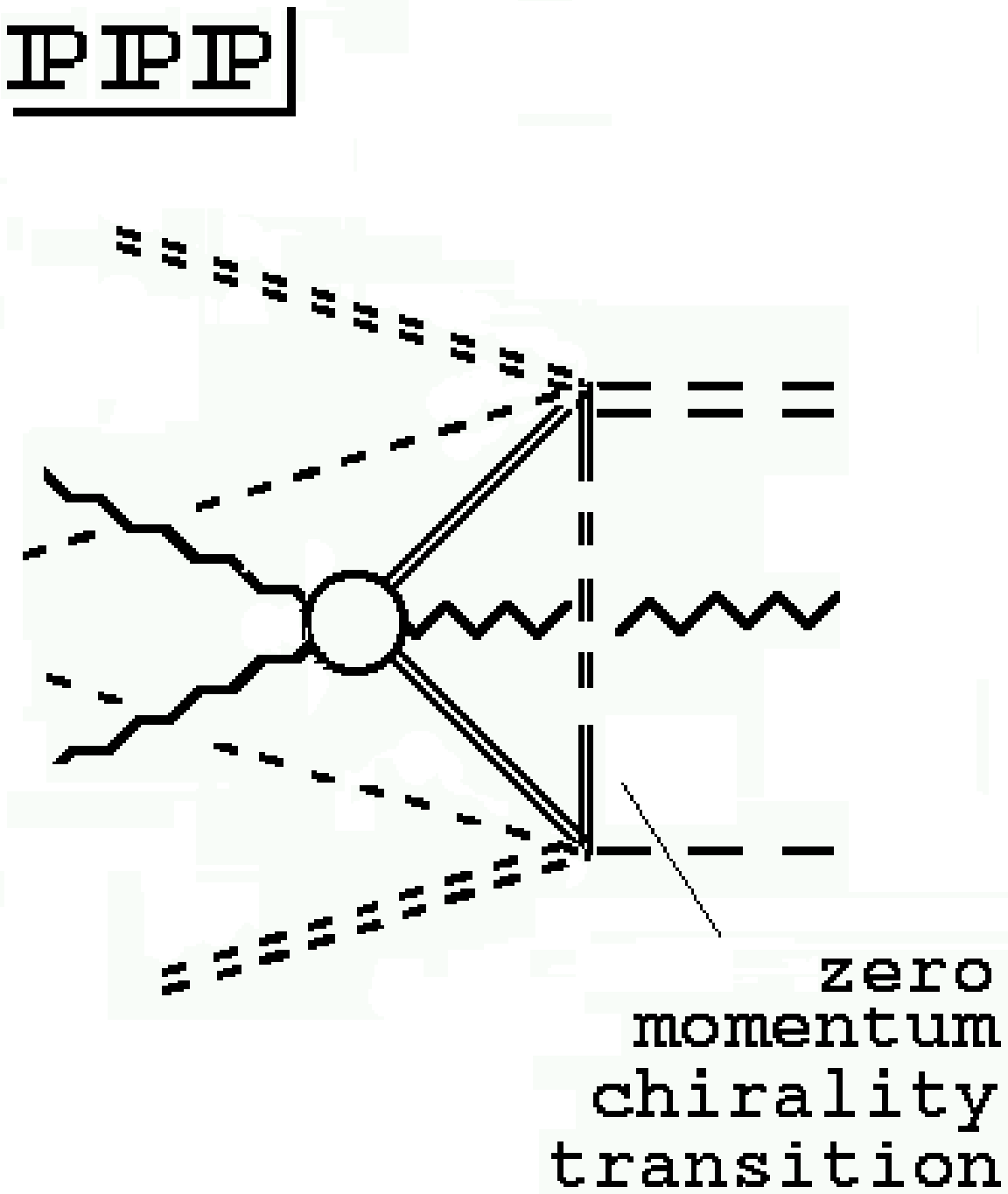}
\hspace{0.2in}
\epsfxsize=4.1in
\epsffile{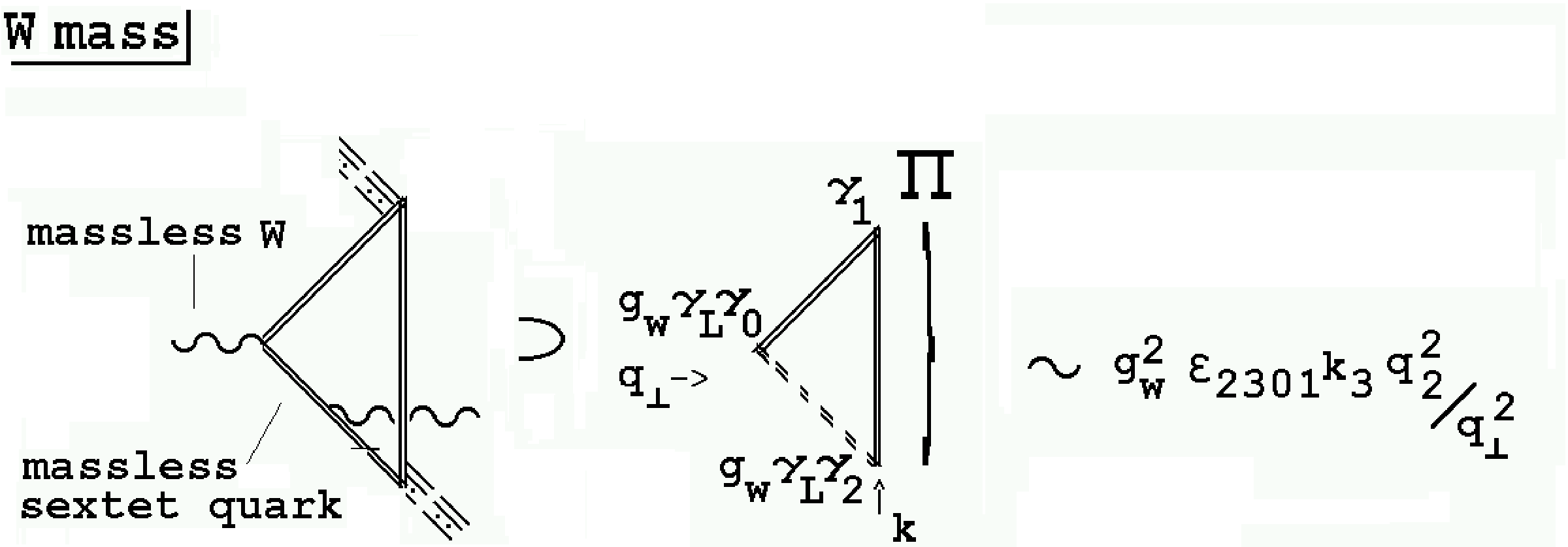}

Figure 6. Anomaly Vertices
\end{center}

In a flavor channel, the presence  
of the chirality transition breaks a chiral symmetry that would otherwise be present
in the massless theory and, in so doing, produces
an anomaly pole that is a Goldstone Boson particle pole. As we will see below, 
this is how multi-regge bound-states are produced. 
We can regard an anomaly pole Goldstone boson as the quark/antiquark state that is  
initially produced within the corresponding
triangle diagram. In this case, 
either the quark or the antiquark has to be in a zero momentum 
``negative energy'' state. Alternatively, we can regard the Goldstone boson
as the produced combination
of a ``normal'' quark/antiquark state together with $k_{\perp} =0$ anomalous gluons.
The ``semi-classical'' gluon field can be viewed as compensating for the chirality
transition of the zero momentum quark or anti-quark. Equivalently, we can say that,
in the process of creating the particle state,
there is a shift of the Dirac sea Fermi surface in that a 
negative energy hole state becomes physical via the introduction of a compensating
``wee gluon'' state.

In the pion/pion/pomeron and triple pomeron vertices 
appearing in Fig.~5, the anomaly pole appears
in a U(1) channel involving only gluons. In this case, the anomaly
pole acts as a $k_{\perp}$ conserving $\delta$-function
that crucially connects scaling divergences in separate channels.
Again we can say that the anomaly pole is created by
a shift of the Fermi surface in which a 
negative energy hole state becomes physical via the introduction of a compensating
``wee gluon'' state.

\mainhead{6. THE SMALL $\beta$-FUNCTION AND COLOR SUPERCONDUCTIVITY}

QUD and massless QCD$_S$ share three, closely related, small $\beta$-function 
properties that are very important
for the construction of multi-regge amplitudes.
\begin{enumerate}
\item{The asymptotic freedom constraint is saturated\cite{kw}.} 
\item{An infra-red fixed-point keeps the $\beta$-function small\cite{bz} 
and also produces reggeon kernels that scale canonically in the $k_{\perp}$ infra-red
region.}
\item{An asymptotically free fundamental representation scalar field can
be used\cite{gw,cel} in both QCD$_S$ - to break
SU(3) color to SU(2) to give CSQCD$_S$ (``color superconducting QCD$_S$''), and in
QUD - to break SU(5) color to SU(4) to give CSQUD (``color superconducting QUD'').}
\end{enumerate}
Because of these three properties, the multi-regge reggeon diagrams of
CSQCD$_S$ and CSQUD can be used to obtain the corresponding 
QCD$_S~$ and QUD amplitudes, respectively. 
(As we will discuss a little more later, a major virtue
of this construction procedure is that the 
anomaly interactions that are introduced, effectively, provide
a resolution of the Gribov ambiguity in light-cone quantization of the unbroken
theory.) The scaling reggeon kernels play an essential role in the occurrence of  
anomaly-coupled infra-red divergences that produce physical amplitudes, while
the absence of a cut-off as the full color group is confined is essential for the
development of the  Critical Pomeron reggeon critical phenomenon.

We will first discuss SU(2) confinement, which is essentially the same in both 
QCD$_S$ and QUD. We will then
specifically discuss massless QCD$_S$ and afterwards turn to QUD. However,
we will see that massless QCD$_S$ contains
a large array of massless Goldstone boson states that probably jeopardize the actual
existence of the S-Matrix. The only possibility (that we know of) to add masses 
to the massless states of QCD$_S$, while preserving the dynamics, 
is to embed the theory in QUD.
According to our arguments, QUD has no massless particles 
and so there should be no threat to the (perhaps unique) existence of the S-Matrix. 

\mainhead{7. SU(2) CONFINEMENT}

In the limit giving SU(2) color confinement, a central role is played by 
color zero ($I=0~~$) combinations of gluon reggeons that we call ``anomalous gluons''.
These are sets of gluon reggeons that carry color charge parity C not equal to their
signature $\tau$. The ``analytic'' definition of signature for a reggeon state
is (for vector reggeons) simply the odd/even number of reggeons. 
For SU(2), if $I=0$, only $\tau=-1$ anomalous combinations are possible.
There is also a ``group-theoretic'' signature\cite{gaw} which has
to coincide with the analytic definition.
When parity (P) is conserved, the combination of incoming and outgoing particle
states to which a reggeon combination couples (via a vertex of the form
of Fig.~6(a)) can be assigned a parity which is also
carried by the reggeon state. The signature 
is the sign given by
a TCP transformation of the complete coupling. Since T simply 
interchanges the ingoing and outgoing particles, it must be that $\tau=CP~$. 
As a result, anomalous gluon couplings must have $P=-1$. 

If we consider forward 
scattering then $P= -1$ for the coupling implies that there must be a parity
change between the initial and the final scattering state. 
In a parity-conserving vector theory, such a change can only come from
an anomaly vertex that contains a zero momentum chirality transition. 
Hence anomalous gluons can only couple via anomaly vertices. 
They can couple to external states via anomaly 
vertices of the form of Fig.~7(a). 
\begin{center}
\epsfxsize=5in
\epsffile{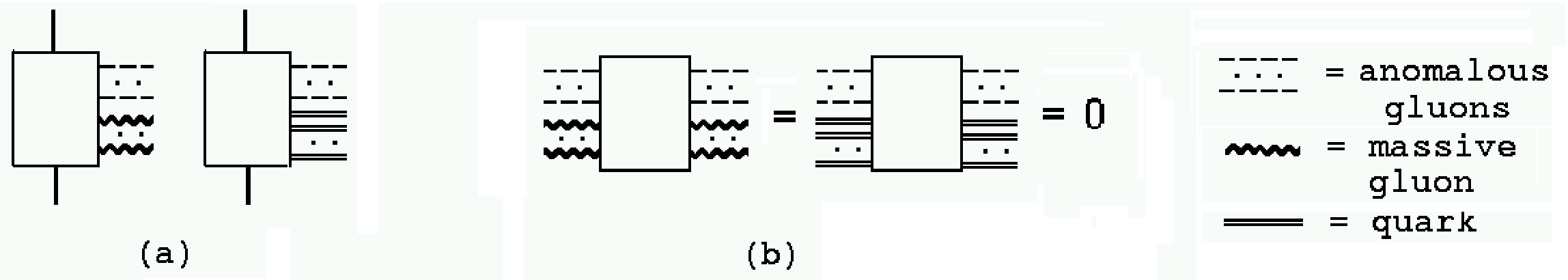}

Figure 7. (a) External Vertices, and (b) Single Channel Vertices.
\end{center}
There will, however, be no anomalous gluon interactions of the form of Fig.~7(b),
because of the absence of anomaly vertices for single channel interactions.

As SU(2) color is confined (with a $k_{\perp}$ cut-off) all divergences
exponentiate, except for the infra-red divergence which occurs\cite{arw02} within
a set of $I=0$ anomalous gluons when the transverse
momenta of all the gluons is scaled uniformly to zero. 
As we discuss again below, because of the scaling property
of the reggeon kernels, this divergence is preserved as the anomalous gluons 
self-interact. Also, there are no interactions of the form of Fig.~7(b) that 
would exponentiate the divergence. Divergences in separate channels can be
coupled\cite{arw02} via anomaly vertices and as a result the
simplest divergent di-triple-regge amplitudes
have the form shown in Fig.~8. 
\begin{center}
$~~~~~~~~~~~~$ \epsfxsize=5in
\epsffile{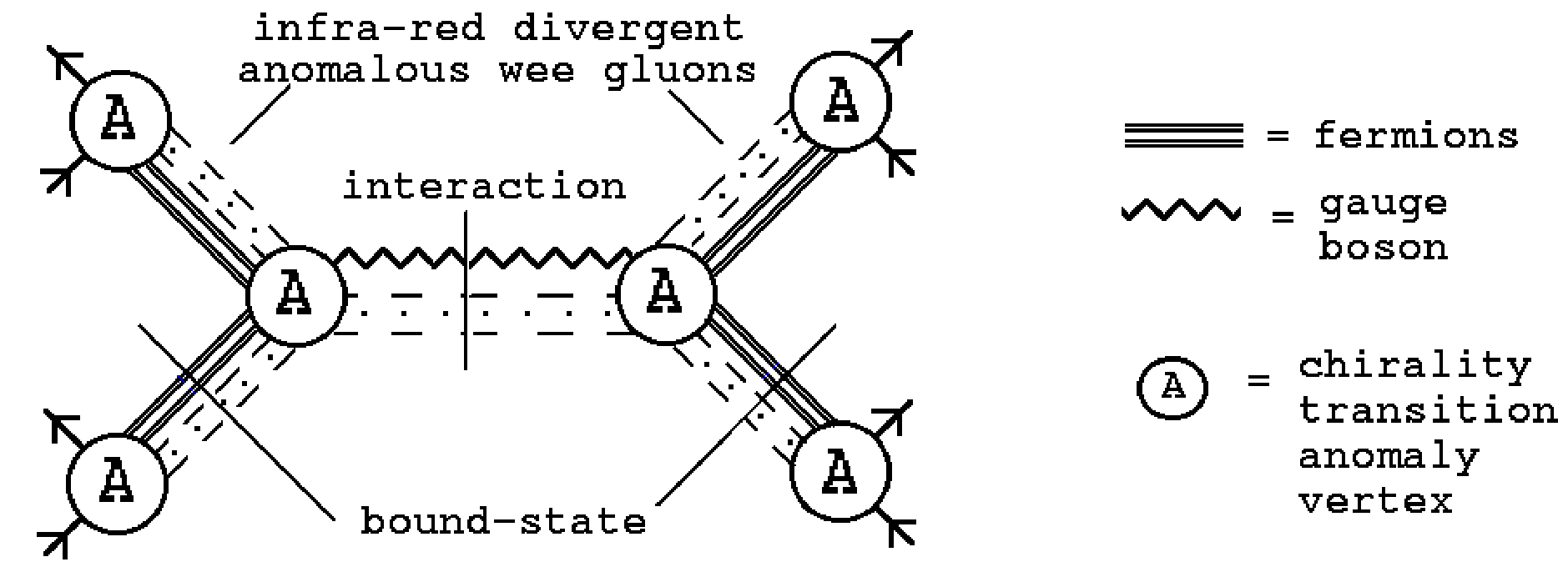}

Figure 8. A Divergent Di-Triple-Regge Amplitude.
\end{center}
When external left-handed massive vector mesons are utilised, as in 
Fig.~4, they provide initial anomalous couplings, that contain Goldstone boson
anomaly poles, as illustrated.
If any of the anomalous gluon configurations in Fig.~8 is replaced by 
non-anomalous $I=0$ gluon reggeons then fermion loop interactions, of the form
of Fig.~9, exist that will (because of the absence of Ward identity 
zeroes - due to the $k_{\perp}$ cut-off)
exponentiate the amplitude to zero.
Consequently, the divergence necessarily involves 
anomalous gluons in every channel.
\begin{center}
\epsfxsize=5.5in
\epsffile{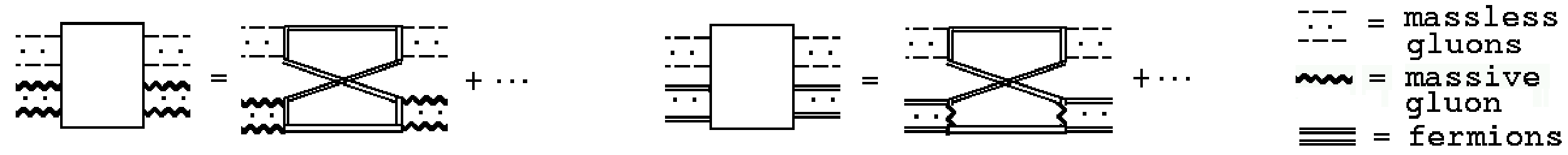}

Figure 9. Fermion Loop Interactions
\end{center}

After the divergence is factored off, 
the remaining bound-state ``physical amplitudes'' 
contain a $k_{\perp}=0~~$ ``wee gluon condensate'' as a relic of the divergence.
In the first approximation, provided by Fig.~8, 
\newline \parbox{5in}{
\begin{itemize}
\item{The bound-states are anomaly poles that appear as $I = 0$ combinations of 
fermions in the anomalous wee gluon background.}
\item {Interactions 
are due to a finite transverse momentum gauge boson, that carries $I=0$, 
in the same wee gluon background.}
\end{itemize}}
\hspace{0.1in} 
\parbox{0.6in}{
\epsfxsize=0.5in
\epsffile{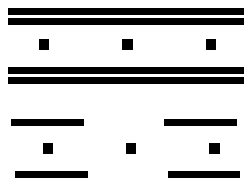}

\vspace{0.2in}

\epsfxsize=0.5in
\epsffile{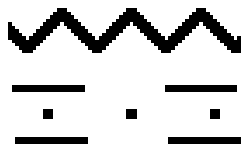}
}
\newline Because the bound-states are anomaly poles, they must be Goldstone bosons and so
there must be a corresponding chiral symmetry - when the confined color is SU(2). 
As we noted above, 
we can regard the Goldstone bosons as initially created by a product of quark/antiquark 
operators  provided we 
remember that removal of the wee gluon component corresponds to a chirality transition.
As we also observed, we can regard the chirality transition 
as equivalent to the zero-momentum 
shift of the Dirac sea Fermi surface during the interaction. Using this language, 
the lowest-order scattering process of Fig.~8 can be described as follows.

An anomaly pole Goldstone Boson ``pion'' is created by the product of a physical
quark field and a zero momentum ``unphysical'' antiquark
field in which the Fermi surface is shifted.
The antiquark becomes physical, via a chirality transition, that introduces
an accompanying ``semiclassical'' anomalous wee gluon field (condensate)
that effectively
moves the Fermi surface back to it's perturbative location. 
In the scattering process, there is a large rapidity perturbative exchange interaction 
in which the wee gluon fields of the incoming pions are transformed
into those of the outgoing pions by anomaly couplings that 
involve further shifts of the Fermi surface. The final state pions are  
created via further shifts of the Dirac sea that reabsorb the anomalous wee 
gluon fields. 

Higher-order effects will add
interactions amongst the gluons in an anomalous gluon state. Divergent 
fermion loop contributions are exponentiated out 
and the remaining gluon interactions
can be described by kernels $K$ 
that, because of the infra-red fixed-point, scale in the infra-red region.
As the kernel interactions are iterated the degree of divergence does not increase.
Instead, in an integral involving a product of many kernels, there is
a distinct contribution from each intermediate state. This divergence 
can be isolated and the remaining integrations factorized as illustrated in Fig.~10. 
\begin{center}
\leavevmode
\epsfxsize=4.5in
\epsffile{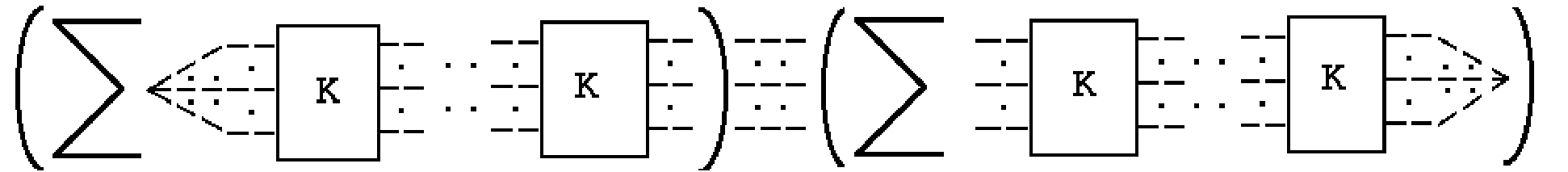}

Figure~10. Factorization of a Wee Gluon Divergence.
\end{center} 
The factorising residues imply that the condensate component of the bound-states 
becomes a ``wee gluon distribution'' in which the gluons have 
zero transverse momentum but carry non-zero ``wee'' longitudinal momenta.  
The wee gluon distribution will have a dynamical
scale, in addition to the cut-off and reggeon masses that we have introduced
as scales. The condensate scale, associated with the divergence, at this point,
is a parameter that has to be determined by matching with supercritical pomeron
theory, as we discuss briefly below.. 

As illustrated in Fig.~11, in higher orders there will also be interactions, 
containing anomalies, between
the anomalous gluon components of the scattering states and the exchanged ``interaction''
state. These interactions are the source of the 
triple pomeron interaction, including
supercritical pomeron interactions, and the mass generation for vector bosons.
\begin{center}
\leavevmode
\epsfxsize=2.3in 
\epsffile{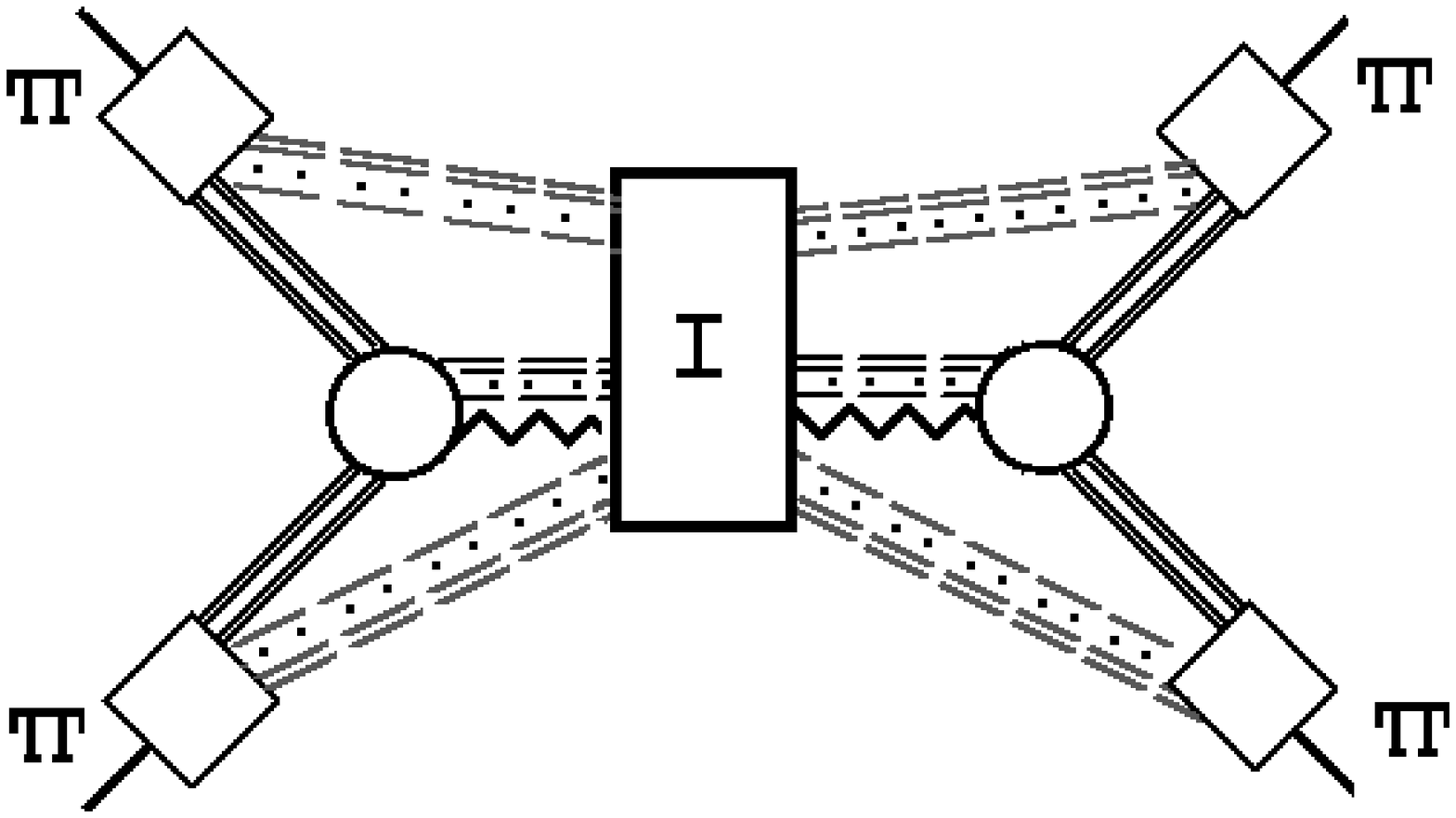}

Figure~11. Interactions Involving Anomalous Gluons in the Scattering States.
\end{center}

\mainhead{8. STATES AND AMPLITUDES IN MASSLESS QCD$_S$}

In CSQCD$_S$, the first approximation to the 
interaction is the even signature combination of a massive SU(2) singlet gluon 
reggeon in the odd signature anomalous wee gluon condensate. As a result, there
is the pomeron/reggeon exchange degeneracy that defines the supercritical
RFT phase. The wee gluon condensate also produces interactions, of the form of
Fig.~11, that include those of the RFT supercritical pomeron. The restoration of 
SU(3) color should then produce the Critical Pomeron 
as the QCD$_S$ high-energy interaction. 
As part of the approach to the phase transition, 
the SU(2) singlet gluon will become massless and 
decouple. Simultaneously, the wee gluon condensate will disappear 
and a corresponding dynamical degree of freedom will appear. That is, the 
shifting of the Dirac sea (the chirality transitions within the anomaly vertices
- the triple pomeron vertex in particular) will become dynamical.
As we already discussed in the Introduction, the shifting of the Dirac sea 
is the ``order parameter'' of
the transition. In the supercritical phase this degree of freedom is ordered into
a single, semi-classical, wee gluon gauge field contribution, 
while in the sub-critical phase it is random. 
For this to happen, longitudinal
vector meson interactions, which at first sight should decouple as the 
color symmetry breaking is removed, must still be present - at zero light-cone
momentum. In fact, the role
of zero light-cone momentum, longitudinal, gluons is a major ambiguity of
light-cone quantization\cite{arw84}.  
By constructing the high-energy behavior of $QCD_S$ via $CSQCD_S$ we 
resolve this ambiguity.

Much remains to be done to map the reggeon diagrams of CSQCD$_S$ onto
the pomeron diagrams of supercritical RFT. In particular, the
parameters of CSQCD$_S$, including the reggeon condensate value, have to be matched
with the RFT parameters. Because the $k_{\perp}$ cut-off is a relevant parameter
for the RFT critical point, it is very important that this can be removed before the 
color symmetry restoration. In \cite{arw93} we describe arguments, that we first
developed a lot earlier, why the number of quark flavors 
should be reflected in the pomeron intercept as a parameter and gave 
general arguments why 
the asymptotic freedom constraint should be saturated, as it is in QCD$_S$. 
We also discussed, 
how the Critical Pomeron scaling functions contain a variety of critical indices
that have to be closely related to properties of the underlying field theory.
In particular, in the zero mass limit, the elastic scaling function collapses to a
single critical index function 
in a way that suggests this index should be a direct property
of an infra-red fixed-point in an underlying vector theory. Very likely, the well-defined
critical indices of the QCD$_S$ fixed-point
are, in fact, the critical indices of the Critical Pomeron scaling functions. In this
sense, this would make the Critical Pomeron the S-Matrix manifestation of the
underlying fixed-point field theory.

The physical states of QCD$_S$ correspond to the Goldstone bosons of CSQCD$_S$. 
Included are
all flavor non-neutral pseudoscalar mesons containing only triplet quarks,
which there will be many of. A potential flavor neutral meson 
mixes with pure gluon states and, hence, does not appear as a Goldstone boson. 
Since quark and antiquark representations are equivalent 
when the gauge symmetry is SU(2), there are also ``nucleon'' Goldstone boson states,
reflecting real chiral symmetries\cite{kog} of $CSQCD_S$.
These states will become baryons by aquiring an additional quark (or
antiquark). To discuss this we need to study more 
the role of the SU(2) singlet quarks in $CSQCD_S$. 
Since these quarks are not Goldstone bosons, they can not be
physical states. They can, nevertheless, appear in regge exchanges.
In particular, there will be a regge 
exchange involving the combination of a
Goldstone boson ``nucleon'' and an SU(2) reggeized quark 
that can become a normal, reggeized, nucleon as SU(3) color is restored.
Understanding this better is important because
the corresponding formation of QUD states is considerably more complicated

There will also be meson and nucleon states involving the two sextet quarks.
If the sextet quarks have the right electroweak quantum numbers and 
the electroweak sector is added, the sextet mesons (``sextet pions'') 
will be eaten by the electroweak vector bosons and the only remaining
sextet states will be the sextet nucleons (apart from the $\eta_6$, which will have
an electroweak scale mass due to mixing with the pomeron sector). Because 
of the larger color factors that are involved, the sextet states will dominate
production cross-sections once the energy is well above the effective threshold
(involving pomeron exchange).

Because there are no corresponding Goldstone bosons in CSQCD$_S$, there will be
no hybrid sextet/triplet quark states. Consequently,
the lightest sextet nucleon will be stable.
The (triplet quark) proton is lighter than the neutron only because 
the current mass of the $d$ quark is bigger than that of the $u$. We expect
effective quark masses to be generated by the embedding of QCD$_S$ in QUD.
However, sextet quark current masses 
must remain zero for sextet pions to combine with the massless electroweak vector bosons
to produce massive states. 
(More strictly, it is the combination of sextet and triplet quarks that couples to
the vector bosons that must have zero current mass.)
Therefore, the sextet nucleon mass difference 
has to be entirely electromagnetic in origin, and   
the $N_6$ will be stable. If the sextet
quark dynamical mass is given by the top quark mass, as discussed earlier,
the $N_6$ mass should be  $\approx  500~ GeV$ and the $P_6$
mass should be just a little higher. Because 
the neutral $N_6$ will not only be stable but will also 
dominate ultra high-energy  cross-sections it, potentially, provides
a natural explanation for both the production and dominance of ``dark matter''.

Since triplet and sextet quarks can not combine to form bound states,
sextet nucleons should not form bound states with triplet nucleons. 
(If pion exchange provides the binding force for nucleons
to form nuclei, 
there is no common ``pion'' to bind sextet and triplet nucleons as nuclei.)
Therefore, we can expect the sextet nucleons to form separate ``dark matter nuclei''.

The QCD$_S$ pomeron produces rising cross-sections via a critical phenomenon
but, nevertheless,   
is a regge pole with RFT interactions and has the factorizing couplings
that are essential if its wee parton component is to be universal and reproduce
``vacuum properties''. This is a, rarely emphasized, essential requirement for
the validity of an infinite momentum parton model\cite{arw84}.
To the extent that the wee gluon component of infinite momentum states is the 
equivalent of a finite energy vacuum, we can say that the ``QCD$_S$ vacuum'' is a 
critical phenomenom of dynamical, zero momentum, fermion chirality transisitions.
The physical states that are stable in this ``vacuum'' are 
much fewer and the interaction much simpler than in conventional QCD.
Moreover, we have a diagrammatic description
of how the states are formed. The spectrum is
consistent with, but much less than, just requiring confinement
and chiral symmetry breaking.
There is also no BFKL pomeron, no odderon, and no glueballs.
In general, there is much better agreement\cite{ce} with experiment!

\mainhead{9. QUD STATES AND AMPLITUDES}

To construct QUD states
and amplitudes we again start with all reggeons carrying global color and with
masses generated by fundamental representation scalars. 
For QUD, in contrast to
QCD$_S$, it is essential that 
the initial fermion masses are generated by condensates. This
has important consequences for the structure of the anomaly interactions, containing
zero momentum chirality transitions, that remain in the massless theory after the
symmetry breaking is removed.
It is, however, the effect of left-handed reggeon couplings
that is the central element in the structure of QUD reggeon diagram divergences. 

Since a left-handed interaction  
violates parity, fermion interactions of the form of Fig.~9 will exist for
both normal and anomalous color parity combinations of 
massless gauge bosons in which one or more of the bosons has 
a left-handed coupling. 
As long as a $k_{\perp}$ cut-off is in
place, therefore, the absence of Ward identity zeroes will 
then imply that these interactions
exponentiate to zero any divergences involving massless left-handed reggeons.
Consequently, in QUD the analog of ``anomalous gluon divergences'' in QCD 
can not involve left-handed gauge boson reggeons.
Anomalous divergences coupled to 
chirality fluctuations, can involve 
only a maximal non-abelian vector subgroup. The resulting
``strong interaction'' pomeron is therefore a singlet under an SU(3) subgroup. 
Because the SU(5) color symmetry of reggeons is a global symmetry, an SU(5) singlet 
pomeron interaction (involving a minimum of four gauge boson reggeons - with three
forming an anomalous configuration)
can be obtained by, effectively, summing over all SU(3) subgroups.
It is then an outcome of
QUD that the strong interaction is produced by a, parity conserving, 
vector interaction that is invariant under an underlying SU(3) gauge group 

To construct QUD amplitudes, 
we use the $SU(3)\otimes SU(2)\otimes U(1)$ breakdown described earlier and
denote the various subgroups of SU(5) as in Figure 12. 
\begin{center} 
\epsfxsize=2.6in
\epsfbox{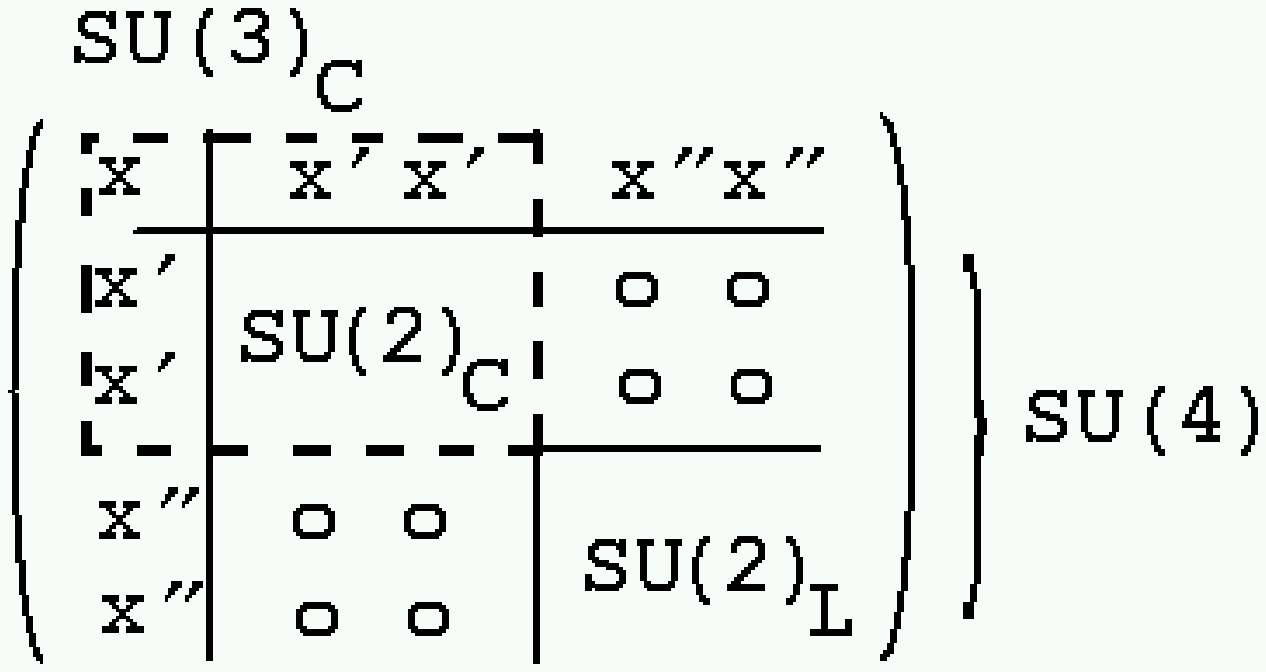}

Figure 12. SU(5) subgroups
\end{center}
The SU(3)$_C$ subgroup
will correspond to a vector interaction. 
We will first restore the SU(2)$_C$ symmetry, then go straight to the SU(4)
symmetry. Restoring the SU(3)$_C$ symmetry will coincide with the final
restoration of the full SU(5) symmetry. The x, x' and x'' vectors will, therefore,
remain massive until the final symmetry restoration. 
We initially consider 
di-triple-regge amplitudes of the form of Fig.~8 with massive, left-handed,
x'' vector bosons as the external states.  

Since SU(2)$_C$ is a vector symmetry, after it is restored   
the states will be chiral Goldstone bosons ($~\pi_C$'s~).
These will be $~qq$, $\bar{q}\bar{q}$, and $q\bar{q}$ pairs 
in a condensate, where the $q$'s are {\bf 3's, 6's,} \& {\bf 8's} under SU(3)$_C$. 
{\bf 8's} are real wrt SU(3)$_C$, but contain complex doublets with respect to 
SU(2)$_C$ that will have a chiral symmetry.

The interactions that are selected by the SU(2) condensate and that 
will also produce SU(3)$_C$ singlets (after the final symmetry
restoration) are just
\begin{enumerate}
\item{a massive x gluon in the condensate - corresponding to the supercritical
pomeron.}
\item{SU(2)$_L \otimes$ U(1) bosons 
in the condensate, corresponding to W$^{\pm,0}$ and Y vector bosons.}
\end{enumerate}
At first sight, the left-handed bosons will have interactions with the 
condensate, of the form
of Fig.~9, that will give exponentiating divergences. However, 
interactions of the form shown in Fig.~11, 
involving the (vacuum) wee gluons of the scattering states,
give left-handed 
$W^{\pm}$ \& $Z^0$ exchanges a mass, via the last anomaly vertex appearing in 
Fig.~6 - that corresponds to mixing with the sextet $\pi_C$'s. 
(The mixing with octet pions will
disappear after the SU(3)$_C$ symmetry is restored.)
Because of this mixing, the massive vector bosons will carry
the sextet SU(2) flavor symmetry in a manner that will allow them to interact via
anomalies only. This eliminates interactions of the form of Fig.~9.

Restoring SU(4) symmetry to obtain CSQUD
involves only left-handed and abelian vector
bosons and so all new divergences exponentiate, leaving only   
states and interactions that are SU(4) invariant.
``Leptons'' are present 
as reggeon bound states of ``elementary leptons'' and ``octet pions''.
The  SU(2)$_L \times$U(1) quantum numbers of octet $\pi$'s are
$ (2,\frac{1}{2}),~ (1,-1)$, and $(3,-1)~$ and so the elementary
lepton component has (modulo gauge boson contributions) 
the generation structure of the Standard Model.

The SU(3)$\times$SU(2)$_L \times$U(1) content of the bound-state leptons is
\begin{enumerate}
\item{ $(e^-,\nu)$ candidate   
\newline $ ~~~~\leftrightarrow ~
(1,2,-\frac{1}{2}) \times (8,1,1)(8,2,-\frac{1}{2})$
$~\leftrightarrow~ SU(5)~ singlet~-~ 45^*\times40\times45^*$} 
\item{$e^+$ candidate 
\newline $~~~~\leftrightarrow ~
(1,3,1) \times (8,2,-\frac{1}{2})(8,2,-\frac{1}{2})$
$~\leftrightarrow~ SU(5)~ singlet~-~ 15\times45^*\times45^*$}
\item{$(\mu^-,\nu)$ candidate
\newline $~~~~ \leftrightarrow ~
(1,2,\frac{1}{2})  (1,2,-\frac{1}{2})  (1,2,-\frac{1}{2}) 
\times (8,1,1)(8,2,-\frac{1}{2})$
\newline $~~~~ \leftrightarrow~ SU(5)~ singlet 
~-~ 5\times45^*\times45^*\times40\times45^*$}
\item{$(\tau^-,\nu)$ candidate 
\newline $~~~~ \leftrightarrow
(1,2,-\frac{3}{2})  (1,2,\frac{1}{2})  (1,2,\frac{1}{2}) 
\times (8,1,1)(8,2,-\frac{1}{2})$
\newline $ ~~~~\leftrightarrow~ SU(5)~ singlet~ -~ 40\times5\times5\times40\times45^*$}
\end{enumerate}

Clearly ``hadron'' states will be present that contain a triplet 
``pion'' and, or, a triplet ``nucleon'' combined with octet quark pions. 
Similarly there will be sextet ``nucleons'' combined with octet pions.
(The sextet pions will have already been absorbed by the vector bosons.) 
When we have described
the fate of octet pions after the full SU(5) restoration, it will be clear why
leptons must involve octet pions. It will not be so clear why SU(5) invariant 
quark states should contain octet pions. It seems likely that this is an outcome of the
SU(2)xU(1) anomaly cancellation in the infra-red, but we have yet to understand this.
It may also be that anomaly cancellation effects of this kind are only evident
after the full symmetry restoration. 

The restoration of SU(5) symmetry is an elaborate phenomenon
that certainly needs much more study and that, as yet I only partially understand.
My current understanding includes the following. 
\begin{itemize}
\item{The pomeron interaction becomes critical - as an SU(3) subgroup interaction
that is summed over subgroups.} 
\item{The $\gamma,W^{\pm}$ and $Z^0$ wee gluon component 
becomes even signature.}
\end{itemize}
The octet quarks, which at first sight might have seemed unwanted, 
are fundamental for the SU(5) invariance and the generation structure of the states.
\begin{itemize}
\item{After SU(3)$_C$ is restored, 
the octet $\pi$'s are no longer Goldstone bosons. Instead, they 
contribute to bound-states as (what would normally be unphysical) anomaly
poles generated by large $k_{\perp}$ quark and antiquark pairs which carry opposite
sign energies. This is a subtle, and very crucial, 
phenomenon that will be elaborated on in much more detail
in \cite{amtm}.}
\item{SU(3)$_C$ reality implies that the octet $\pi$'s have no anomaly coupling to the 
pomeron and so leptons have no strong interaction and no infra-red SU(3)$_C$ 
mass generation.}
\item{With the octet $\pi$'s contributing to states only at large $k_{\perp}$,
the SU(2)$_L\otimes$U(1) symmetry appears in low $k_{\perp}$ interactions
via the SU(2) flavor symmetry of the sextet sector. This is another phenomenon 
that will be discussed in more 
detail in \cite{amtm}.}
\item{The SU(2)$\otimes$U(1) quantum numbers of the octet $\pi$'s
implies that low $k_{\perp}$ states will have the singlet/doublet
structure of the Standard model.}
\item{Since the octet pions are embedded in leptons at large $k_{\perp}$, we expect
(although we have yet to develop a proper argument) that the bound-state 
lepton contribution
to the infra-red SU(2)$_L\otimes$U(1) anomaly is equal to the perturbative
contribution. This would imply the existence of three generations of leptons.}
\item{The SU(2)$_L\otimes$U(1) anomaly cancellation then
requires the formation of three generations of quark hadrons that similarly
contain the octet pions.}
\end{itemize}

Although we will not attempt to discuss it in any detail here, the mass spectrum will be 
generated by a combination of factors. Firstly, there will be straightforward
perturbative reggeization effects that carry, initially, the momentum scale corresponding
to the evolution scale of the underlying gauge theory. Large color factors and the,
related, high mass sector will emphasize the SU(3) strong interaction. In addition,
anomaly interactions, analagous to that generating the $W$ and $Z$ masses, will mix 
all the reggeon states. There seems every reason to believe that the large
disparity in (generalised) color factors can produce a wide range of scales within
the S-Matrix. The most obvious example being the disparity in scales 
for the strong and electroweak interaction produced by Casimir scaling of SU(3) color 
factors. Without a better understanding of the anomaly interactions
and the related wee gluon distributions it is unclear how many parameters could be 
involved, in general. In particular, 
although CP violation could easily be a consequence of
chiral anomaly dominance of interactions, at this point it is obviously not 
clear whether it is actually necessary. Nevertheless, to suggest  
that the physical mass spectrum could emerge from QUD is clearly not unreasonable.

\mainhead{10. GENERAL PROPERTIES}

Apart from explaining why the strong interaction is an SU(3) vector theory 
and why it is that the left-handed interaction aquires a mass,  
there are many other general features of QUD that are encouraging.
\begin{enumerate}
\item{
The experimentally attractive SU(5) value of the Weinberg angle should hold -
even though there is no proton decay !}
\item{Small $\alpha_u$ should be the explanation of small neutrino masses.}
\item{The QCD sector agrees better\cite{ce} with experiment (than conventional QCD)
- no glueballs, no BFKL pomeron and no odderon.}
\item{The existence and dominance of Dark Matter is naturally explained.}
\item{The electroweak symmetry breaking shares key features with successful\cite{ffrs} 
``Walking Technicolor'' theories that are 
consistent with precision constraints.}
\item{The high mass QCD sector produces unification without supersymmetry.}
\item{There are no unwanted symmetries constraining the mass spectrum.}
\item{QUD is contained in a single SO(10) representation - the {\bf 144}.
Although there is no S-Matrix for the enlarged
representation (according to our arguments) this could be
relevant for \{string?\} unification with gravity?} 
\end{enumerate}

\mainhead{11. WHAT SHOULD BE SEEN AT THE LHC?}

In this Section I will briefly review what should be seen at the LHC and
how, in particular, double pomeron processes can provide the definitive
proof that a sextet sector has appeared. A more extended discussion can be found in
\cite{arw05}.

Because large cross-sections are involved, the emergence of the 
sextet sector should be obvious. 
The immediate evidence will be that multiple vector boson and jet x-sections are 
much, much, larger than expected. $<p_{\perp}>$ should undergo a major increase 
from the low energy hadron scale and move significantly towards the electroweak scale.
There will, however, be competing explanations for these effects, such as 
black holes, sphalerons, etc..

A priori, $N_6\bar{N}_6$ pair production (dark matter) should be seen - with 
the $N_6$ mass, perhaps, $\sim 500 ~GeV$. Unfortunately, this will be difficult to detect,
since missing energies of several hundred GeV will be common.
Also the low energy $N_6$ hadronic 
cross-section, for collisions in a calorimeter, is probably small.
$P_6\bar{P}_6$  pair production should be seen
- assuming the $P_6$ is not too unstable. Again, however,
a massive charged particle with a large production x-section
will not be immediately identified with the sextet sector ! 

Surprisingly, perhaps, the double pomeron cross-section could actually provide 
the most definitive evidence for the existence of the sextet sector.
With the pomerons detected via Roman pots, the environment is clean and 
well controlled. $W$ and $Z$ pairs will be produced in the double pomeron 
cross-section via sextet pion anomaly poles. 
As (triplet quark) pion pairs dominate the double pomeron
cross-section at low mass, so $W$ and $Z$ pair production will dominate the
cross-section at the electroweak mass scale. Naively, a factor of
$~[ F_{\pi_6}/ F_{\pi_3}]^4~
~ \centerunder{\raisebox{0.5mm}{${\scriptstyle >}$}}{${\scriptstyle \sim}$}
~O(10^{12})~$ is involved in relating
sextet and triplet sector ``pion'' anomaly-pole cross-sections. However,
this is not very useful since normal double-pomeron production of pions 
does not involve vector states and so does not proceed via anomaly poles.

When $|k_{\perp}|$ is electroweak scale,
the double-pomeron $ W$ and $Z$ pair amplitude for producing jets is comparable
with a standard jet amplitude that has, apart from anomaly
loops that are O(1), the same propagators and couplings. This suggests
that the jet cross-section from double-pomeron $W$ and $Z$ pairs will be
comparable with the non-diffractive jet
cross-sections predicted by standard QCD.
While the $~\pom~ W^+W^-~\pom~$ and $~\pom~ Z^0Z^0~\pom~$
vertices should vary only slowly with $k_{\perp}$, the $pp~\pom~$ 
vertices have strong $ k_{\perp}$-dependence. This implies there should be
an extremely large x-section at small $t$. 

In the initial low luminosity running, an ``extremely large x-section''
could be detected by TOTEM in combination with the CMS central detector 
(assuming it is operational) - where it should be
straightforward to look for the leptonic decays of $W$ and $Z$ pairs. Apart from
factors of sextet isospin, the $Z$ pair cross-section will have the same order
of magnitude as the $W$ pair cross-section. Consequently, 
some spectacular events should be expected,  
in which protons are tagged and only (a multitude of)
large $E_T$ charged leptons are seen in the central detector.

FP420 (with Roman pots designed to look for a Standard Model Higgs' boson) 
will take over during the high luminosity running 
and should surely see an enhanced cross-section, even if it is too small 
to have been seen by CMS/TOTEM. With the planned parameters for FP420, 
the $W$ and $Z$ pair cross-section should overwhelm all other physics.

The observation of a very large double-pomeron cross-section for $W$ and $Z$ pairs
would imply that the longitudinal components of the $W$ and the $Z$  
have direct strong interactions. The only known possibility for this
is the existence of the sextet sector and, as we have discussed, to give a well-defined 
theory this sector has to be embedded in QUD !

After the combination of $\pom$, $W/Z$, and jet physics has established that 
sextet quark physics is definitively discovered, 
the search for ``dark matter'' will become all important.
The cross-section for double-pomeron production
of stable $N_6\bar{N}_6$ pairs (with a pair mass
$~ \centerunder{\raisebox{0.5mm}{${\scriptstyle >}$}}{${\scriptstyle \sim}$}
~1~TeV$) 
could be large enough that it will be definitively seen by the forward pot
experiments. It will be a spectacular process to look for - via the following.
\begin{enumerate}
\item {The tagged protons determine
a very massive state is produced.}
\item {No charged particles are seen
in any of the detectors.}
\item{Having low energy, the $N_6$ hadronic 
cross-section will, probably, be small
but some hadronic activity may be seen in the central calorimeter.}
\item {Charged lepton comparison would 
allow a separation wrt the multiple $Z^0$ production of neutrinos.}
\end{enumerate}
Of course, if the $P_6$ is relatively stable, 
and not too different in mass from the $N_6$, it would be much simpler
to first detect $P_6\bar{P}_6$ pairs.

\mainhead{12. COMMENTS AND PERSPECTIVE}

Although there is not as much speculation in what I have described 
as probably appears to be the case to the non-specialist reader (almost everyone),
there certainly is a considerable amount. Even with the publication of \cite{amtm},
much will remain that needs to be both better established and also 
better understood. Many  
questions will remain unanswered and many details will still be missing. 
As a result, it will surely be some time
before serious calculational procedures can be developed. 
Nevertheless, I am hopeful that my discovery of QUD, and all of it's remarkable 
properties, will eventually demonstrate that (contrary to popular belief) 
solving the infra-red problem of constructing 
physical states that produce a unitary S-Matrix may actually 
be more difficult, more special, 
and ultimately at least as fundamental as the
solving of the ultra-violet problem of a field theory.

While it may appear that I have introduced
way too much that is radical, and in conflict with the current theoretical 
paradigm, I would argue that this was not by choice. I have been led along the
path I have followed by logical necessity. Most signicantly, I have been led 
to, what seems to me at least to be, a very beautiful proposition, 
that the relevant entity for particle physics
is the bound-state S-Matrix of a very special, small $\beta$-function, 
massless field theory that, at first sight, is an ``unparticle theory''\cite{hg}. 
In fact, as I have described, the zero momentum ``Fermi surface'' 
of the massless Dirac 
sea offers crucial possibilities for wee gluon interactions via anomalies that prevent
the scale invariance property of the unparticle theory from carrying over into
the physical S-Matrix (although understanding and elaborating how 
all the S-Matrix scales originate is a significant
part of the work that remains to be done). 
More importantly, perhaps, it seems that the massless gauge
theory need only be evident in (and, therefore, need only exist as a quantum field theory 
in) short-distance perturbation theory. Mass generation becomes an S-Matrix
property which is, effectively, separated from the problem of having a sufficiently
well-defined short-distance field theory. 

It is important to emphasize that, besides my construction 
via multi-regge theory, there is no other formalism capable of 
constructing bound-state scattering amplitudes. Without this ability
it would not have been possible to envision the existence of an S-Matrix within
a field theory with the properties of QUD. 
In effect, I diagrammatically construct the 
high-energy S-Matrix via infra-red and ultra-violet cut-off manipulations 
that determine the contribution of fermion anomalies
in the (multi-regge region) perturbation expansion. Although I have very little 
idea as to how the finite energy S-Matrix might be analagously obtained, it 
seems unlikely that the states I find could appear as intermediate states
in any off-shell Green's functions. (As we noted in the Introduction, if there were 
a connection to field operator Green's functions, infra-red scale
invariance would be in conflict with \cite{asv} the existence of a particle spectrum.)
The zero momentum chirality transitions and resulting anomalous wee gluon interactions
are introduced via the formation of asymptotic states and so are 
clearly particular to the S-Matrix. As a result, I anticipate that the S-Matrix is the 
only well-defined non-perturbative element of the theory. 
Although this is a radical notion, according to current thinking, it 
is well-established historically that it is 
fully viable from a practical (experimental) viewpoint\cite{hs}. It could also 
have the great advantage that (as a matter of principle) there would be no
need to confront the overwhelmingly difficult, and so far elusively intractable, 
problem\cite{jw} 
of constructing a full, non-perturbative, quantum field theory (with or without
a mass gap) in four dimensions. 
 
We can ask, of course, why the massless field theory has to be QUD. My answer
has been that I demand the appearance of the Critical Pomeron. However, I can also
phrase this requirement 
in terms that would be, perhaps, more familiar to the general reader
via comments already made in previous Sections. 
The (infra-red fixed-point) small $\beta$-function is required, firstly 
for the persistence of the 
scaling wee gluon interactions that enhance infra-red fermion anomaly 
interactions, and
secondly to allow the color-superconductivity starting point that resolves the
quantization ambiguities associated with Gribov copies and Gauss's law. 
The vector interaction non-abelian gauge group has to be as large as 
SU(3) to produce, via scaling wee gluon interactions, a
universal wee gluon distribution that can carry vacuum properties. This property
is surely essential\cite{arw84} for the existence of an infinite momentum 
``parton model'' that allows asymptotically free perturbation theory to produce an
ultra-violet finite S-Matrix. If the vector gauge group is larger than SU(3), 
the anomalous wee gluon scaling interactions are more complicated and
the universal wee parton property is lost. The SU(3) gauge group can, however, be   
extended by left-handed interactions, that aquire a mass via the anomalies, 
since the only effect
is to also generate bound state masses. This is a beneficial effect in that it
alleviates potential S-Matrix infra-red problems. (Essentially,
the infinite momentum, wee parton, properties are not affected.) Asking that this
extension generates masses for all bound-states while introducing no short-distance
anomaly then brings us close to, if not directly to, QUD. Therefore, 
I believe that although I have funneled my discussion 
through the Critical Pomeron, in fact all of the properties needed to obtain a
well-defined particle S-Matrix come together to uniquely select QUD. 

It is currently accepted, almost without question, that ``non-perturbative'' QCD 
and all similar
unbroken non-abelian gauge theories should be
well-defined by the euclidean path integral. This is taken to imply that
there must be a physical S-Matrix and that, moreover, the physical states appear as
intermediate states in off-shell Green's functions (derived from the path 
integral) of appropriate operators. Although there is no evidence to support
this hypothesis, the considerable 
phenomenological success of various ``non-perturbative QCD'' formalisms, particularly
lattice QCD, implies there must be some approximate truth in the assumptions.
Nevertheless, at the level where we are asking to understand why a theory is uniquely 
chosen by nature it is important to emphasize that approximations are being made
and that there are significant assumptions involved.

The existence of a short-distance field theory may be essential, not only for  
the large momentum finiteness of the S-Matrix, but also, as discussed in \cite{arw00}, 
for local analyticity properties. 
That the S-Matrix can be obtained from ``non-perturbative'' off-shell Green's functions 
does not, however, appear to be essential for any of it's basic properties. The
global analyticity domains that are normally thought to be a consequence of an 
off-shell field theory probably follow from the construction of physical high-energy 
amplitudes via the perturbation expansion. In fact, when the fields are massless and 
bound states related to infra-red anomalies are involved there is probably no 
general reason to expect a connection between Green's functions and the S-Matrix. 

In general, there is not even a formal
property of a non-abelian gauge theory path integral which implies that a 
unitary, bound-state, S-Matrix can be derived via Green's functions.
Even worse, because of infra-red problems, the path integral itself is, most likely,
not well-defined - both because of the infinite
volume convergence
problem in four dimensions and, more seriously perhaps, 
because of the ambiguity of the function space implied by the Gribov copy problem. 
Since there are no ``non-perturbative'' methods for constructing gauge theory
S-Matrix amplitudes that do not, effectively, appeal to the formal euclidean 
functional integral, 
to seriously discuss whether a unitary S-Matrix exists in a general gauge theory 
is a highly non-trivial problem. 

Of course, it would be incredible 
if the Standard Model, with all of it's 
complexity, has the underlying simplicity that I have suggested. Nevertheless,
all the necessary ingredients are present and if the predicted effects of the sextet sector
are seen at the LHC, I doubt that 
the radical/heretical nature of what I am proposing will impede the rapid rise of 
interest in QUD that will surely ensue. It is important to emphasize that, in principle,
there is no freedom for variation in QUD. It is an ``all or nothing'' explanation
of the origin of the Standard Model. Although my current understanding has not allowed
any really quantative predictions, an unavoidable  
central element and most striking prediction of QUD is 
that the new physics producing electroweak symmetry breaking is due to an additional 
strong interaction sector that will be abundantly evident at the LHC.


\begin{thebibliography}{99}

\bibitem{cri}  A.~A.~Migdal, A.~M.~Polyakov and K.~A.~Ter-Martirosoyan, 
{\it Zh. Eksp. Teor.  Fiz.} {\bf 67}, 84 (1974); 
H.~D.~I.~Abarbanel and J.~B.~Bronzan, {\it Phys. Rev.} {\bf D9}, 2397 (1974).

\bibitem{amtm} A.~R.~White, ``A Massless Theory of Matter'' - to appear.

\bibitem{kw} K.~Kang and A.~R.~White, {\it Int. J. Mod. Phys.} {\bf A2},
409 (1987).

\bibitem{arw05} A.~R.~White, {\it Phys. Rev.} {\bf D72}, 036007 (2005). 

\bibitem{wm} W.~J.~Marciano, {\it Phys. Rev.} {\bf D21}, 2425 (1980).

\bibitem{bww}~E.~Braaten, A.~R.~White and C.~R.~Willcox, {\it Int. J. Mod. 
Phys.}, {\bf A1}, 693 (1986).

\bibitem{hg} H.~Georgi, {\it Phys. Rev. Lett.} {\bf 98:221601}, (2007).

\bibitem{asv} A.~Armoni, M.~Shifman and G.~Veneziano, hep-th/0403071.

\bibitem{nik} See S.~I.~ Nikolsky, {\it Phys. Atom. Nucl.} {\bf 62},
2048 (1999), for references.  

\bibitem{cores} Z.~Cao, L.~K.~Ding, Q.~Q.~Zhu, Y.~D.~He,
{\it Phys. Rev.} {\bf D56} 7361 (1997).

\bibitem{CDFb} CDF Collaboration, ``CDF-QCD Group Run 2 Results''
- http://www-cdf.fnal.gov/physics/new/qcd/inclusive/index.html. This is
an early Run 2 analysis which clearly makes my point. A variety of 
``more sophisticated'' jet algorithms have since been developed that improve 
the agreement between theory and experiment.

\bibitem{CDF} CDF Collaboration (T. Affolder et al.), 
{\it Phys. Rev. Lett.} {\bf 88} 042001 (2002).

\bibitem{mm} M.~Moshe, {\it Phys. Repts.} {\bf 37}, 256 (1978).

\bibitem{arw84} A.~R.~White, {\it Phys. Rev. } {\bf D29}, 1435 (1984).
Although I did not properly understand the anomaly-based dynamics at the time, 
this paper already describes why the Critical Pomeron is needed to provide
an infinite-momentum parton model and also how it is 
that QCD$_S$ gives the Critical Pomeron.

\bibitem{arw98} A.~R.~White, {\it Phys. Rev.} {\bf D58}, 074008 (1998). 
This paper describes all the necessary multi-regge limits, as well as the 
construction of multi-reggeon diagrams via reggeon unitarity. See also
A.~R.~White, {\it Int. J. Mod. Phys.} {\bf A11}, 1859 (1991),
H.~P.~Stapp H P, in {\it Structural Analysis of Collision Amplitudes},
North Holland (1976) and A.~R.~White, ibid. 

\bibitem{arw00} A.~R.~White, ``The Past and Future of S-Matrix Theory'',
published in ``Scattering'', edited
by E.~R.~Pike and P.~Sabatier (Academic Press, 2002).

\bibitem{bs} J.~B.~Bronzan and R.~L.~Sugar, {\it Phys. Rev.} {\bf D17}, 
585 (1978), this paper explicitly organizes the transverse momentum diagram results from 
H.~Cheng and C.~Y.~Lo, Phys. Rev. {\bf D13}, 1131 (1976), 
{\bf D15}, 2959 (1977), into reggeon diagrams satisfying reggeon unitarity. 

\bibitem{fs} E.~Fradkin and S.~H.~ Shenker, Phys. Rev. {\bf D19}, 3682 (1979);
T.~Banks and E.~Rabinovici, Nucl. Phys. {\bf B160}, 349 (1979).

\bibitem{fl} V.~S.~Fadin and L.~N.~Lipatov, {\it Nucl. Phys.} {\bf B406},
{\it Nucl. Phys.} {\bf B477}, 767 (1996) and further references therein.

\bibitem{arw02} A.~R.~White, {\it Phys. Rev.} {\bf D66}, 056007 (2002).

\bibitem{bz} T.~Banks and A.~Zaks, {\it Nucl. Phys.} {\bf B196}, 189 (1982).

\bibitem{gw} D.~J.~Gross and F.~Wilczek, Phys. Rev. {\bf D8}, 3633 (1973).

\bibitem{cel} T.~P.~Cheng, E.~Eichten and L.~F.~Li, Phys. Rev. {\bf D9},
2259 (1974).

\bibitem{gaw} P.~Goddard and A.~R.~White, {\it Nucl. Phys.} {\bf B17}, 45 (1970).

\bibitem{arw93}  A.~R.~White, {\it J. Mod. Phys.} {\bf A8}, 4755 (1993).

\bibitem{kog} J.~B.~ Kogut, M.~A.~Stephanov, D.~Toublan, 
J.~J.~M.~ Verbaarschot and A.~Zhitnitsky, {\it Nucl. Phys.} {\bf B582},
477 (2000).

\bibitem{ce} It is well-known that conventional QCD anticipates a 
wide variety of glueball states and that, so far, none have been convincingly observed -
E.~Klempt, {\it Int. J. Mod. Phys.} {\bf A21}, 739 (2006).
Similarly the BFKL pomeron and the Odderon are perturbative components 
of conventional QCD that have not been definitively observed -
C.~Ewerz, ``The Perturbative Pomeron and the Odderon: Where 
Can We Find Them?'' hep-ph/0403051.

\bibitem{ffrs} R.~Foadi, M.~T.~Frandsen, T.~A.~Ryttov, F.~Sannino, arXiv:0706.1696 [hep-ph].

\bibitem{hs} H.~P.~ Stapp, {\it Phys. Rev.} {\bf D3}, 1303 (1971).

\bibitem{jw} ``Quantum Yang-Mills Theory'',
A.~M.~Jaffe, E.~Witten, Clay Mathematics Institute Millenium Prize Problem (2000).

\end{thebibliography}
\end{document}